\documentclass[useAMS,usenatbib]{mn2e} 
\usepackage{graphicx}
\usepackage{subfigure}
\usepackage{epsfig}
\usepackage{times}
\usepackage{natbib}
\usepackage{amsfonts}
\usepackage{amsmath}
\usepackage{amsbsy}

\newif\ifAMStwofonts

\def\Djdot{$\Delta \dot{J}/\dot{J}_{\rm cons}$}

\def\mavg{$\dot{m}$}
\def\m0mc{$\dot{m}/\dot{m}_{\rm c}$}
\def\rin{$r_{\rm in}$}

\def\rc{$r_{\rm c}$}
\def\del2r{$\Delta r_2/r_{\rm in}$}
\def\delr{$\Delta r/r_{\rm in}$}

\newcommand{\bhl}{}
\newcommand{\ehl}{}
\newcommand{\bhm}{}
\newcommand{\ehm}{}

\newcommand{\beq}{\begin{equation}}
\newcommand{\eeq}{\end{equation}}

\title{Accretion Discs Trapped Near Corotation}
\author[C.\ R.\ D'Angelo \& H.\ C.\ Spruit]
       {Caroline R.\ D'Angelo and
         Hendrik C.\ Spruit \\  Max Planck Institute for
         Astrophysics, Garching, Germany} 
\pagerange{\pageref{firstpage}--\pageref{lastpage}}
\pubyear{2011}

\begin{document}

\maketitle
\label{firstpage}

\begin{abstract}
\bhm  We show that discs accreting onto the magnetosphere of a rotating
  star can end up in a ‘trapped’ state, in which the inner edge of
  the disc stays near the corotation radius, even at low and varying
  accretion rates. The accretion in these trapped states can be steady
  or cyclic; we explore these states over wide range of parameter
  space. We find two distinct regions of instability, one related to
  the buildup and release of mass in the disk outside corotation, the
  other to mass storage within the transition region near
  corotation. With a set of calculations over long time scales we show
  how trapped states evolve from both nonaccreting and fully accreting
  initial conditions, and also calculate the effects of cyclic
  accretion on the spin evolution of the star. Observations of cycles
  such as found here would provide important clues on the physics of
  magnetospheric accretion. Recent observations of cyclic and other
  unusual variability in T Tauri stars (EXors) and X-ray binaries are
  discussed in this context. \ehm
\end{abstract}

\section{Introduction}
Many accreting stars show evidence of the effects of a strong stellar
magnetic field regulating the accretion flow onto the star. Accreting
X-ray pulsars, for example, show flux modulation on timescales between
$\sim 10^{-3}-10^{3}$ seconds, which is attributed to the magnetic
pole sweeping through our line of sight, illuminated by matter
accreting along field lines \citep{1973ApJ...179..585D}. In some
pulsars, this probe of the star's period has shown an evolution in the
spin period of the star in time (e.g. \citealt{1997ApJS..113..367B}),
which is attributed to the interaction with the surrounding
material. On the other end of the energy scale, T~Tauri stars also
often have strong magnetic fields (up to 1-2 kG), and show some
evidence that their period is regulated by the disc-field interaction
\citep{2008ApJ...688..437G}.

From the earliest studies of stellar accretion researchers recognized
that a strong magnetic field could substantially influence an
accretion disc surrounding the star. \cite{1972A&A....21....1P}
estimated the `magnetospheric radius' of the disc -- the radius at
which the magnetic field would truncate the disc -- by equating the
ram pressure of free-falling gas to the magnetic pressure of the
field. The inner radius of the disc is thus determined by the magnetic
field strength (assumed dipolar with a magnetic moment $\mu\equiv
B_{\rm S}r^3_*$), stellar mass and the accretion rate onto the star: $
r^{7/2}_{\rm m} \propto \mu^2M^{-1/2}_*\dot{m}^{-1}$. The infalling
material will then add its angular momentum to the star, causing it to
spin faster. \cite{1975A&A....39..185I} noted that if the
magnetospheric radius extends beyond the corotation radius ($r_{\rm c}
\equiv (GM_*)^{1/3} \Omega_*^{-3/2}$; where the Keplerian frequency of
the disc equals the star's rotation frequency), the stellar magnetic
field will spin faster than the gas at the inner edge of the disc,
generally called the `propeller' regime. A standard view expressed in
the literature is that in this case the mass in the disc is expelled
from the system (see below) and the star is spun down. \bhm The exact
relationship between the spin evolution of the star and the accretion
rate will depend on the physics of the disc-magnetosphere interaction
when the inner edge of the disc is close to \rc.\ehm


\bhm Although observations of magnetic accreting stars generally
support this general picture for magnetospheric accretion, questions
remain.\ehm\ A number of persistent X-ray pulsars show spin-down as
well as spin-up (e.g. \citealt{1997ApJS..113..367B}), or a rate of
spin-up much lower than expected based on the accretion rate.  In
addition, many sources show hysteresis, where the luminosity \bhl\
differs between spin-up and spin-down phases \ehl
\citep{2010ApJ...708.1500C}, or even anti-correlates with spin-down
\citep{1997ApJ...481L.101C}. Finally, in persistent sources that show
both spin-up and spin-down, the magnitude of $|\dot{\nu}|$ often stays
nearly constant when the torque changes sign, which is not naturally
explained in a model where the torque scales as a power of \mavg.

More recently, two transient X-ray pulsars were observed to undergo
brief weak outbursts (lasting about 6-7 days) followed by a short
period of quiescence
\citep{2010ApJ...714..894H,2010arXiv1006.1908H}. The outburst
recurrence time (on the order of a month in both sources) is too short
to be naturally explained by the ionization instability model
(e.g. \citealt{2001NewAR..45..449L}), and \cite{2010arXiv1006.1908H}
have suggested (based on the total outburst luminosity) that a
significant amount of mass could remain stored in the disc when the
accretion rate onto the star drops by at least 3-4 orders of
magnitude. This explanation is inconsistent with the standard relation
between accretion rate and inner radius, in which $r_{\rm in} \propto
\dot{m}^{-2/7}$, so that a change of $10^4$ in accretion rate
corresponds to an increase of $\sim 10\times$ in \rin.

In one of these sources, NGC 6440 X-2, \cite{2010ATel.2672....1P}
recently reported the detection of a strong QPO at 1Hz. A similar QPO
was previously detected in SAX J1808.4-3658
\citep{2009ApJ...707.1296P}. This QPO period is of the order
$10^{-2}-10^{-1}$ times the viscous timescale at the corotation radius
\rc\ and hints further at interaction between the disc and the field.

Similar outburst time scales, in units of the viscous time scale, are
seen in a class of young stars called `EXors'. These stars typically
show episodic changes in luminosity (of between 2-5 magnitudes) on
timescales of a few years. The variability timescale implies accretion
rate variability in the inner regions of the accretion disc, where it
interacts with the magnetic field.

Together, these puzzling observations suggest that significant physics
is missing in current interpretations of the phenomena seen in
accreting magnetospheric systems.

\subsection{Accretion at a centrifugal barrier}
\bhl One of the things missing in the standard interpretation is the
insight that mass prevented from accreting by the presence of a
`centrifugal barrier' does not necessarily have to leave the system at
all. In other words, mass transferred from a binary companion onto a
spinning magnetosphere does not have to be `propellered' out.

This was already noted by \cite{1977PAZh....3..262S} and
\cite{1993ApJ...402..593S}. Without mass loss, accretion on the
magnetosphere can instead be {\em cyclic}, with periods of
accumulation outside the corotation radius alternating with accreting
phases. The problem of disc responding to the torque exerted at its
inner edge by a spinning magnetosphere has a well-defined solution
within the standard thin viscous disc formalism without mass loss.
Mass loss in a magnetically driven wind may happen as well, but this
bit of (still poorly known) physics is separate from the effect of a
centrifugal barrier on a viscous disc.

We have studied this problem in \cite{2010MNRAS.406.1208D} and
\cite{2011MNRAS.416..893D} (hereafter DS10 and DS11
respectively). There we classified the properties of time dependent
solutions of the thin disc diffusion equation with a magnetic torque
acting at its inner edge (DS10), and studied the long term evolution
of discs with such torques, including the spindown of the star
(DS10). \ehl

Since the magnetic torques transport angular momentum outward, the
density profile of the disc is altered substantially. The disc remains
truncated only slightly outside \rc\ even when the accretion rate
declines by several orders of magnitude. We found that the formation
of a dead disc often results in the instability described by
\cite{1993ApJ...402..593S}, which we suggested could be operating in
EXors as well as SAX J1808.4-3658 and NGC 6440 X-2.  In the
instability, the disc is initially truncated outside \rc\ and
accretion is suppressed, causing material to build up in the inner
regions of the disc. Eventually, the surface density in the disc
becomes high enough to overcome the centrifugal barrier and accrete
onto the star. Once the reservoir is emptied, the disc again moves
outside \rc\ and the cycle begins again.

In DS10 we concluded that the presence of the instability depended on
the mean accretion rate, \mavg\ and the details of the disc-field
interaction.\bhm We found that the uncertainties in the physics of the
disc-field interaction could be parameterized into two independent,
unknown (but constrained) length scales: the radial width of the region
of interaction, $\Delta r$, and the length scale over which the disc
moves from an accreting to a non-accreting state $\Delta r_2$. We made
a preliminary investigation of the unstable solutions for different
values of our independent parameters and found the shape and duration
of the cycles of accretion varied widely. In the present work we
investigate the parameter space more thoroughly in order to
characterize the instability better and link it with
observations. \ehm

\subsection{Trapped discs}

The modes of accretion of a viscous disc on a rotating magnetosphere
can be classified in three states, all of which have a different
appearance and effect on the star's evolution. We explicitly neglect
the possibility of outflows, so that the disc never crosses into the
propeller regime for any \mavg.

The disc state depends on the ratio of the mean accretion rate through
the disc (\mavg) to the accretion rate that puts \rin\ at \rc
($\dot{m}_{\rm c}$):
\begin{enumerate}
\item{ $\dot{m} \gg \dot{m}_{\rm c}$: \rin$<$\rc, star spins up}
\bhl\item{ $\dot{m} \simeq \dot{m}_{\rm c}$: \rin$\approx$\rc, star spins 
either up or down, inner edge stays near corotation while $\dot m$ varies.}
\item{$\dot{m} \sim 0$: \rin$>$\rc, star spins down, negligible accretion (dead disc)}.
\end{enumerate}
\bhl State (ii) can be further divided into cases where accretion
takes place in a continuous way, and cases where accretion is cyclic,
with bursts of accretion alternating with quiescent phases. Our main
interest in the following paper is this second case of cyclic
accretion. If there is a long-term accretion rate imposed at the outer
edge of the disc, as in a mass-transferring binary, dead disc states
can occur as part of the cycle, as suggested by
\cite{1977PAZh....3..262S}\ehl.

\bhl In\ DS11\ehl, we turned our attention to the long-term evolution
of a viscous disc, and its effect on the \bhl spin\ehl\ evolution of
the star. This \bhl added\ehl\ a new parameter to the problem, the
moment of inertia $I_*$ of the star, and introduced a new
characteristic timescale, $T_{\rm SD}$, the spin-down timescale \bhl\
of the star\ehl.

\bhl When initial conditions are chosen such that \rin\ is initially
inside \rc, but moves gradually outward, we found that the evolution
of the disc often gets `trapped' with \rin\ near corotation, with a
slowly decreasing accretion rate. The same happens when \rin\ is
initially outside \rc, but spindown of the star causes \rc\ to move
out. When it catches up, \rin\ continues to hover slightly outside
corotation as the two move outward. The outward angular momentum
transport due to the magnetic torque is accompanied by low-level
accretion onto the star. We called this phenomenon a `trapped
disc'. It is the intermediate state ii) mentioned above.

The disc does not stay trapped in all cases, however: \rin\ can also
evolve well beyond \rc\ to become a `dead disc' -- the state iii)
above. Whether or not the disc becomes trapped depends n the details
of the disc-field interaction and on the ratio $T_{\rm visc}/T_{\rm
  SD}$ (where $T_{\rm visc}$ is the viscous accretion timescale of the
disc at \rc).

\subsection{Observations and Spin Evolution of Stars with Trapped Discs}

\bhm Accretion in the trapped state can be continuous or cyclic. In
DS11 we limited ourselves mostly to parameter regimes that lead to
continuous accretion, since the short time scale of the cycles makes
it difficult to follow the long-term evolution of the disc. In this
paper we study the cyclic case of trapped discs in more detail, with
less emphasis on the system's long-term evolution. We study in
particular the modification to the expected relationship between
spin-down torque and luminosity, and the appearance of cyclic
accretion.\ehm

We explore the nature and interpretation of these cycles in more
detail here and study how the characteristics of the instability (such
as its period and amplitude) change with the model parameters. These
are discussed in section \ref{sec:cycles}. When accretion is cyclic,
the torque between disc and magnetosphere varies over a cycle; its
average over a cycle is what determines the net spindown or spinup
torque on the star. In sec. \ref{sec:cycle_jdot} we study how this net
torque differs from the steadily accreting case.  Before that, in
section \ref{sec:spin} we discuss the observability of the quiescent
``trapped disc'', and the spin evolution of the star as a function of
accretion rate.

In Sec. \ref{sec:sources} we discuss our results in the context of the
various observations mentioned above (persistent X-ray pulsars, the
short recurrent outbursts in X-ray pulsars, and the episodic accretion
bursts seen in both X-ray pulsars and young stars). \bhl We argue
there\ehl\ that trapped discs may be related to a number of currently
unexplained phenomena seen in magnetospherically accreting sources.

\section{The model for magnetospheric accretion}
\label{sec:model}
We briefly summarize the characteristics of our model for
magnetospheric accretion, reviewing our description of the disc-field
interaction and how this alters the structure of the disc when $r_{\rm
 in} > r_{\rm c}$. 
\bhl For a compact description see DS11; for more detail and our
numerical implementation see\ehl\ DS10. 

As mentioned above, the interaction between an accretion disc and
magnetic field is likely confined to the innermost regions of the
disc, so that most of the disc is shielded from the field. The
coupling between the disc and the field distorts the field lines by
differential rotation, which generates a toroidal field component and
exerts a magnetic torque on the disc. \bhl In a very short time\ehl, the
field lines become sufficiently distorted that they inflate and open, which can
temporarily sever the connection between the field and disc, \bhl before
reconnection events reestablish connection with the star. 
As a result, the variations in the magnetic field associated with the
magnetosphere interaction\ehl\ will take place on timescales of order
the rotation period of the star, $P_*$
\citep{1985A&A...143...19A,1996ApJ...468L..37H,1997ApJ...489..890M,1997ApJ...489..199G};
much shorter than the cycle times, which take place on a variety of
viscous time scales in the disc.

It is\ehl\ sufficient to adopt a time-averaged value for the strength
of the generated toroidal field component, $\eta \equiv B_\phi/B_z$
and the width of the interaction region \delr. The size of the
interaction region is unknown, so it is a free parameter in our model;
we assume only that $\Delta r/r < 1$, and explore values in the range
suggested by numerical simulations. (\bhm For a more extensive
discussion of time variability in $\eta$ and $\Delta r$ see DS10.\ehm)

When the inner edge of the disc \rin\ is outside corotation \rc, a
magnetic torque $T_B$ acts at the inner edge. The condition that this
torque is transmitted outward by the viscous stress requires: \beq
3\pi\nu\Sigma(r_{\rm in})r^2\Omega_{\rm K}(r_{\rm in})=T_B.
\label{vt}
\eeq This yields a boundary condition on the surface density at
\rin. When \rin\ is inside \rc\ however, the viscous torque and the
surface density at \rin\ vanish: the standard case of a viscous disc
accreting on a slowly rotating object applies. The viscous torque at
\rin\ thus varies rapidly over the transition width $\Delta r$ around
\rc. This can be taken into account in (\ref{vt}) by making $T_B$ a
function of \rin, 
\beq T_B=y_\Sigma(r_{\rm in})T_0,
\label{tb}
\eeq where $y_\Sigma$ is a suitably steep function with the property
that it varies from 0 to 1 over the width $\Delta r$:
$y_\Sigma\rightarrow 0$ for $r_{\rm c}- r_{\rm in}\gg \Delta r$,
$y_\Sigma\rightarrow 1$ for $r_{\rm in}-r_{\rm c}\gg \Delta r$. For
$T_0$ we approximate the field strength at \rin\ as that of the star's
dipole component, so the magnetic torque exerted by the field is 
\beq
T_0=\eta\, \mu^2\Delta r/r^4,
\label{t0}
\eeq 
where $\mu$ is the star's dipole moment and $\eta$ a coefficient of
order unity. For numerical reasons, it is important that $y_\Sigma$ be
a smooth function; we use a $1+\tanh$-profile of width $\Delta
r$. For the viscosity we use a fixed radial dependence, $\nu=k_\nu
r^{1/2}$, corresponding to an $\alpha$-disc of constant $\alpha$ and
aspect ratio $H/r$.\ehl

When $r_{\rm in} < r_{\rm c}$ the disc can accrete onto the star. The
position of the inner edge of the disc \bhl is then related to
the\ehl\ accretion rate. We use a standard estimate
\citep{1993ApJ...402..593S}, \bhl which equates the magnetic torque
with the torque needed to keep the accreting mass corotating with the
star. It can be written as
\begin{equation}
\label{mdota}
\dot m_{\rm a} = \frac{\eta \mu^2}{4\Omega_*r_{\rm in}^5}
\end{equation}
(the subscript $_{\rm a}$ stands for `only in the case of accretion')
and holds only for \rin$<$\rc. For \rin$>$\rc, the accretion rate
drops to zero due to the centrifugal barrier. The relation between
\rin\ and $\dot m$ thus also changes steeply around \rc. This is
another critical element that needs to be taken into account. We
incorporate it by writing: \beq
\label{eq:dotm}
\dot m_{\rm co}(r_{\rm in}) =y_m(r_{\rm in})\dot m_{\rm a}(r_{\rm in})
\eeq where $y_m(r_{\rm in})$ describes the steep variation in the
transition zone, and $\dot m_{\rm co}$ is the mass flux {\em measured
  in frame comoving with the inner edge of the disc}: \beq \dot m_{\rm
  co}=\dot m+2\pi r_{\rm in}\Sigma(r_{\rm in})\dot r_{\rm
  in},\label{dotmco} \eeq where $\dot r_{\rm in}$ is the rate of
change of the inner edge radius (the distinction between $\dot m$ and
$\dot m_{\rm co}$ was a crucial point missing in the formulation of
ST93). For the `connecting function' $y_m$, with limiting values 1
inside corotation and 0 outside, we again use a smooth $ \tanh$-
function. The width of this transition must be of the same order as
the transition width $\Delta r$ introduced above, but the physics
determining it is rather different, so we keep it as a separate
parameter $\Delta r_2$, and explore it independently.\bhm By
introducing $\Delta r$ and $\Delta r_2$ as the transition length
scales from one boundary condition to other, we reduce the
uncertainties of the transition region to two free but constrained
parameters, which we vary systematically to explore the range of
unstable solutions.\ehm

Eq. (\ref{mdota}) and (\ref{dotmco}) define an equation for \rin($t$).
Together with the standard thin disc diffusion equation, we have a set
of equations, second order in time and space, for $\Sigma$ and \rin.
The equations are stiff inside corotation since one of the time
derivatives (for \rin) disappears. A numerical method appropriate for
this circumstance has to be used.

The steady-state case, $\partial_t=0$, is illustrative. The solution of the thin disc
diffusion equation can be written as
\beq
\label{eq:sig}
3\pi\nu\Sigma ={T_B\over\Omega(r_{\rm in})r_{\rm in}^2}\left({r_{\rm in}\over r}
\right)^{1/2}+\dot{m}\left[1-\left(\frac{r_{\rm in}}{r}\right)^{1/2}\right],
\eeq
where $T_B$ is given by (\ref{tb}) and $\dot m$ by (\ref{mdota}, \ref{dotmco}).

For \rin\ well outside corotation (\rin\ - \rc $\gg \Delta r$), $\dot
m$ vanishes and we have a dead disc. The surface density is then
determined by the first term on the RHS. The steady outward flux of
angular momentum in this case has to be taken up by a sink at some
larger distance, otherwise the disc could not be stationary as
assumed. This sink can be the orbital angular momentum of a companion
star, or the disc can be approximated as infinite. The latter is a
good approximation for changes in the inner regions of the disc, if we
consider short time scales compared with the viscous evolution of the
outer disc.

\subsection{Characteristic numbers}
\label{sec:standard}

We choose \rc\ (the star's initial co-rotation radius) for our
length scale, and define $t_{\rm c} \equiv \nu/r^2_{\rm c}$, the
viscous accretion timescale at \rc\, as our characteristic timescale.
We adopt $\nu_0 \equiv \alpha \left(H/r\right)^2 = 10^{-3}$
for the dimensionless viscosity parameter, and $\eta =0.1$ as the relative 
size of the $B_\phi$ component. We also define a characteristic accretion 
rate $\dot{m}_{\rm c}$:
\begin{equation}
\dot{m}_{\rm c} \equiv \frac{\eta\mu^2}{4\Omega_*r^5_{\rm c}},
\end{equation}
\bhl This is the accretion rate in (\ref{mdota}) that would put the magnetospheric 
radius at $r_{\rm c}$. 

For a typical ms-pulsar\ehl:
\begin{eqnarray}
\dot m_{\rm c} =10^{-14}M_\odot \mathrm{yr}^{-1}\left(\frac{M_*}{1.4M_\odot}\right)^{-5/3}
\left(\frac{B_s}{10^{8}\rm{G}}\right)^2\\
\nonumber
\left(\frac{R_*}{10^6 {\rm cm}}\right)^6\left(\frac{P_*}{100~\rm{ms}}\right)^{-7/3},
\end{eqnarray}
while for the case of a T~Tauri star disc,
\begin{eqnarray}
\dot{m}_{\rm c}= 1.4\times10^{-9}M_\odot \mathrm{yr}^{-1}\left(\frac{M_*}{0.5M_\odot}\right)^{-5/3}\\
\nonumber
\left(\frac{B_s}{1500\rm{G}}\right)^{2}\left(\frac{R_*}{2 R_{\odot}}\right)^{6}\left(\frac{P_*}{7~\rm{d}}\right)^{-7/3}.
\end{eqnarray}
\bhl To characterize the spin change\ehl\ of the star, we
adopt a characteristic torque based on the definition for
$\dot{m}_{\rm c}$:
\begin{equation}
 \dot{J}_{\rm c} \equiv \dot{m}_{\rm c}(GM_*r_{\rm c})^{1/2},
\end{equation}
which is the rate at which angular momentum is added by mass accreting
at $\dot{m}_{\rm c}$. This definition is somewhat arbitrary, but has
the advantage of being independent of \delr\ and \del2r.

\section{Spin evolution and physical properties of a trapped disc}
\label{sec:spin}
\subsection{Observability of trapped discs}
\label{sec:appearance}
The density structure of a trapped disc deviates from the standard
accreting disc. In a dead disc (the limiting case of a trapped disc),
the accretion rate is zero and the surface density profile is
determined by the rate of angular momentum transport from the star to
the disc. 

\bhl Provided that the thin disc approximation remains a good
description for the disc, meaning that the energy generated by viscous
turbulence is dissipated locally the temperature of the dead disc 
can be estimated \citep{1977PAZh....3..262S} from the viscous dissipation. 


The local energy dissipation rate in a viscous disc is:
\begin{equation}
Q^+ = \nu\Sigma(r\Omega'_{\rm K})^2\,=\, {9\over 4}\nu\Sigma\Omega_{\rm K}^2.
\end{equation}
Near the inner edge of the disc where most of the energy is dissipated, 
the disc can be approximated as steady, so (\ref{eq:sig}) applies. For a dead 
disc ($\dot m=0$ and \rin\ well outside corotation), $T_B=T_0$, and the surface density 
as a function of $r$ is given by \ehl
\beq
\label{eq:sdead}
\nu\Sigma= {1\over 3\pi}\left({T_0\over\Omega r^2}\right)_{r=r_{\rm i}}\left({r_{\rm i}\over r}\right)^{1/2}.
\eeq
If this energy is radiated locally as a blackbody, then
\begin{equation}
Q^+=Q^-=2\sigma_{\rm }T^4_{\rm s},
\end{equation}
where $T_{\rm s}$ is the surface temperature of the disc as a function of
radius \bhl (the factor 2 arising because the energy is lost from the two
surfaces of the disc), and $\sigma_{\rm B}$ the Stefan-Boltzmann constant. 
With (\ref{t0},\ref{eq:sdead}) this yields
\begin{equation}
\label{Ts}
\sigma_{\rm B}T_{\rm s}^4= {3\over 8\pi}\eta\mu^2\left({\Delta r\over\Omega 
r^6}\right)_{r=r_{\rm i}}{GM\over r^3}\left({r_{\rm i}\over r}\right)^{1/2},
\end{equation}
which varies with $r$ as $r^{-7/2}$, showing that most of the energy is radiated away
near the inner edge by the mass piled up at the `centrifugal barrier'. If $\nu\sim r^{1/2}$
as we assume in the calculations to follow, (\ref{Ts}) implies $T_{\rm s}\sim r^{-1}$\ehl.

With the nominal object parameters used in \ref{sec:standard}, 
a dead disc \bhl with its inner edge near corotation\ehl\ 
will have a maximum temperature, at \rin, of $T_{\rm s} \approx 300K$ in
 the case of a T~Tauri star and $T_{\rm s} \approx 20000$ K for a ms X-ray pulsar.
 \bhl This estimate, however, only includes the internal energy
 dissipation in the disc.  Both in protostars and X-ray binaries,
 reprocessing of radiation from the central object by the disc usually
 dominates over internal dissipation, especially at larger distances
 from the centre. Nonetheless, it might be possible to identify a disc
 as belonging to the dead class with more detailed information on the
 spectral energy distribution.

\subsection{Accretion rate and angular momentum exchange with the star}
\label{sec:J_mdot}
The interaction between the magnetic field and the disc outside \rc\
removes angular momentum from the star, while the gas accreting onto
the star spins it up. The net torque on the star is thus a function of
both the average accretion rate through the disc and the location of \rin. It is given by:
\begin{equation}
\label{eq:Jdot}
\dot{J} = \dot{m}(GM_*r_{\rm in})^{1/2} - \frac{\eta\mu^2}{r^3_{\rm
 in}}\frac{\Delta r}{r_{\rm in}}y_\Sigma,
\end{equation}
where the first term gives the spin-up from accretion and the second
term is the spin-down from the disc-field interaction. (The spin-down
torque vanishes for $r_{\rm in} < r_{\rm c}$, which we impose with the
$y_{\Sigma}$ smoothing function.) In sec. \ref{sec:model} we defined
the relationship between \mavg\ and \rin, so that $\dot{J}$ will just
be a function of \mavg, \delr\ and \del2r. In the rest of the
section we illustrate the implications of this relationship for the
spin-evolution of the star.

\begin{figure}
  \rotatebox{270}{\resizebox{!}{90mm}{\includegraphics[width=\linewidth]{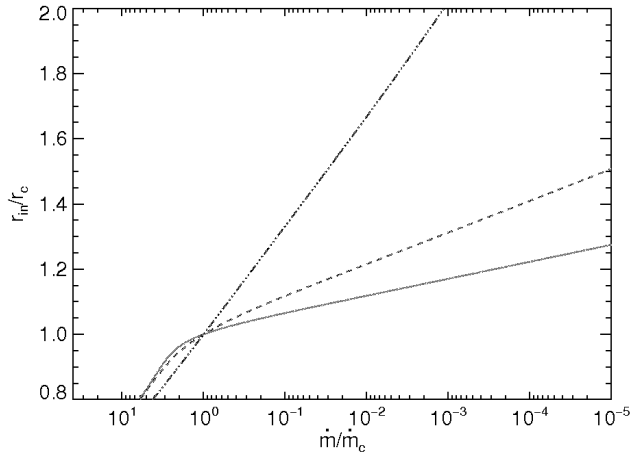}}}
  \caption{\label{fig:rin_mdot} Change in \rin\ as a function of
    \mavg\ for different \del2r. Solid curve: $\Delta r_2/r_{\rm in} =
    0.05$, Dashed curve: $\Delta r_2/r_{\rm in} =
    0.1$,Triple-dot-dashed curve: $\Delta r_2/r_{\rm in} = 0.5$}
\end{figure} 

Fig.~\ref{fig:rin_mdot} shows the relationship between \rin\ and
\mavg\ as \del2r\ changes, to demonstrate how the disc can become
trapped. The three curves show \rin\ as a function of \mavg\
for $\Delta r_2/r_{\rm in} = 0.05$ (solid blue), $0.1$ (dashed green)
and $0.5$ (triple-dot dashed red). At high accretion rates ($r_{\rm
 in} < r_{\rm c}$), all three curves scale as $r_{\rm in} \propto
\dot{m}^{-1/5}$. However as \mavg\ decreases the solutions quickly
diverge. For small \del2r (\del2r\ =$[0.05,~0.1]$), there is a knee in
the solution around $r_{\rm in} = r_{\rm c}$, and \rin\ increases much
more slowly as \mavg\ decreases, so that \rin\ remains close to \rc\
even when \mavg\ has decreased by several orders of magnitude. For
$\Delta r_2/r_{\rm in} = 0.5$, the knee straightens out since the
transition from accreting to non-accreting solutions is much more
gradual. 


\begin{figure}
  \rotatebox{270}{\resizebox{!}{90mm}{\includegraphics[width=\linewidth]{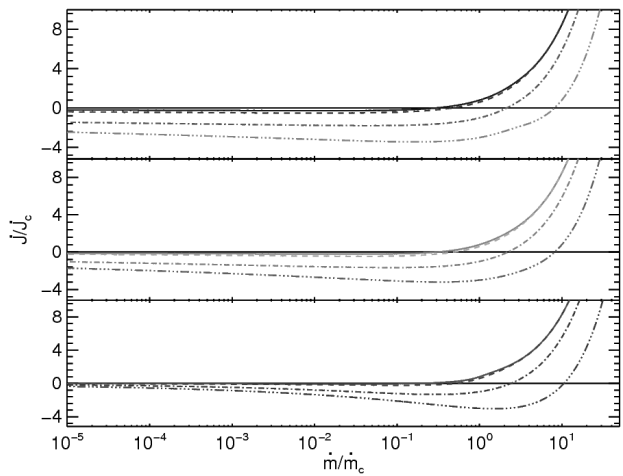}}}
\caption{Change in net torque on the star as a function of \mavg\ for
 different values of \delr\ and $\Delta r_2/r_{\rm
 in}$. From top to bottom, the individual panels correspond to
 $\Delta r_2/r_{\rm in} = [0.05,0.1,0.5]$. The different patterned
 lines correspond to different values of \delr: 0.05
 (solid), 0.1 (dashed), 0.5 (dot-dashed), and 1.0 (triple-dot
 dashed).\label{fig:nu_mdot}}
\end{figure}

Fig.~\ref{fig:nu_mdot} \bhl shows the dependence of the net torque 
$\dot{J}$ on the star on accretion rate \mavg\ehl, for different values of $[\Delta r,
\Delta r_2]$. 
At high accretion rates, the disc is strongly accreting, star spins up
with the torque $\dot{J} \propto \dot{m}^{9/10}$. At lower accretion
rates, the interaction region moves outside \rc, causing the net
angular momentum exchange to change sign, and spinning the star
down. \bhl As the accretion rate decreases further and \rin\ moves 
increasingly far outside \rc\ into the effectively `dead' state, the 
spindown torque decreases because of the decline of the star's
field strength with distance. 

The detailed shape of the curves in figs.
\ref{fig:rin_mdot},\ref{fig:nu_mdot} reflects the ($\tanh$-) shape of
the transition functions $y_\Sigma$ and $y_m$ we have used\ehl. More
generally, however, when $\dot{J} < 0$, the amount of spin down is
directly proportional to \delr, and \del2r determines how far away
from \rc\ the disc can move as \mavg\ decreases.

\section{Cyclic accretion}
\label{sec:cycles}
The possibility of identifying cycles like those seen in DS10 in
actual observations is a interesting prospect. Cycles would provide a
direct observational connection with the trapped disc state phenomenon
that we have identified here as a likely consequence of the
disc-magnetosphere interaction. The shape of the cycles and
circumstances of their occurrence would provide important clues about
the details of the interaction region, which we have simply
parametrized with the two transition widths \delr (for the
torque) and \del2r (for the mass flux)\ehl.

In this section we expand the analysis in DS10 to investigate
the properties of the instability more quantitatively. In
Sec. \ref{sec:cycle_appearance}, we investigate the presence of
the instability as a function of the system parameters. In
sec. \ref{sec:cycle_period} we show how the period and amplitude 
\bhl vary with\ehl\ \mavg, \delr\ and \del2r. In sec. \ref{sec:cycle_jdot} 
we \bhl investigate\ehl\ how the presence of cycles
will change the spin evolution of the star in comparison to a steadily
accreting disc. Finally, in sec. \ref{sec:cycle_transient} we 
examine the appearance of cycles for non-steady accretion in
systems in which the star's spin is evolving.

\subsection{Parameter map of the instability}
\label{sec:cycle_appearance}
In DS10 we showed that the disc instability depends on \delr, \del2r\
and \mavg, and ran simulations in the parameter spaces $[\Delta r,
\dot{m}]$ and $[\Delta r_2, \dot{m}]$ to determine when the
instability occurred.  \bhl Additional\ehl\ simulations of trapped
discs at very low accretion rates show the instability is present for
a larger parameter space than we investigated in DS10, so here we have
repeated our analysis on a larger parameter space and better-sampled grid.

\bhl As in DS10, we keep the rotation rate of the star fixed, since the 
time scale of the cycles is much shorter than the time scale of spin 
changes of the star. The corotation radius is then a fixed distance, and 
defines a unit of mass flux, $\dot m_{\rm c}$ given in sec. \ref{sec:standard}. 
As unit of length we can use \rc, and as unit of time the viscous time scale
$r_{\rm c}^2/\nu(r_{\rm c})$. Apart from a parameter specifying the radial 
dependence of $\nu$, which we keep fixed throughout, the problem is 
then defined by the three dimensionless parameters $\Delta r/r_{\rm in}$, 
$\Delta r_2/r_{\rm in}$, and $\dot m/\dot m_{\rm c}$\ehl. 

\bhl It turns out (discussed below) that this 3-dimensional parameter
space contains 2 nearly separate regions of instability, which can be
characterized using a few 2-dimensional slices instead of having to
scan the entire space\ehl.

In comparison to DS10, the present results survey a larger parameter
range at higher resolution for all three variables. The individual
simulations at higher spatial and time resolution. In total we ran
1545 simulations: 855 in \bhl a 2-dimensional slice\ehl\ $[\Delta
r,\dot{m}]$, and 690 in \bhl a $[\Delta r_2,\dot{m}]$ slice\ehl.  Our
goal was to determine the regions in parameter space where cyclic
behaviour occurs, as well as how the characteristics of the cycles
(such as period, amplitude and appearance) change as a function of the
parameters.

\begin{figure}
      \rotatebox{90}{\resizebox{!}{90mm}{\includegraphics[width=\linewidth]{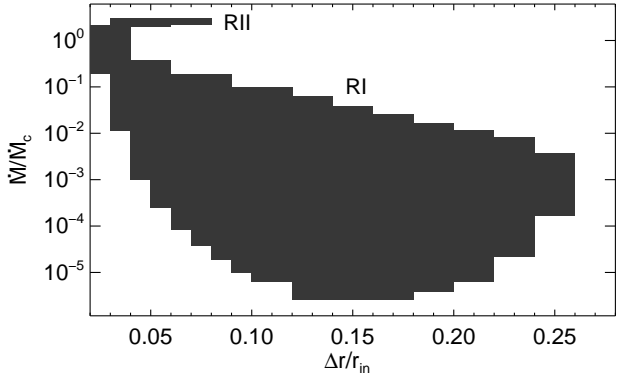}}}
\caption{Instability map in $[\dot{m}, \Delta r/r_{\rm in}]$, keeping
 $\Delta r_2/r_{\rm in} = 0.014$ fixed. The shaded regions denote
 unstable simulations. The instability occurs in two \bhl nearly 
 disconnected\ehl\ regions of the parameter space.\bhl In this 
 slice, one (RI) is present over a large range in $\Delta
 r/r_{\rm in}$ and \mavg, the other in a
 small region in \delr\ at $\dot{m} \simeq
 \dot{m}_{\rm c}$ (RII)\ehl.
 \label{fig:unstable_deltar}}
\end{figure} 

\begin{figure}
      \rotatebox{270}{\resizebox{!}{90mm}{\includegraphics[width=\linewidth]{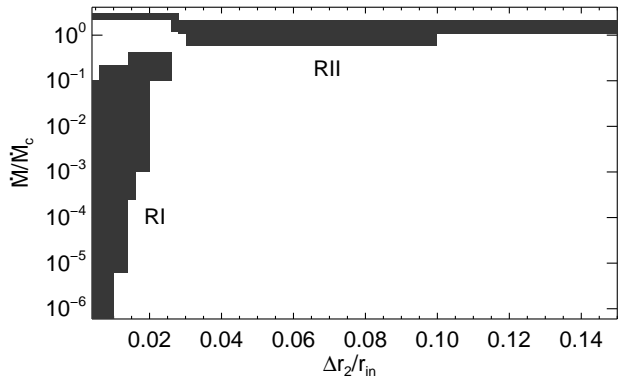}}}
\caption{Instability map in $[\dot{m}, \Delta r_2/r_{\rm in}]$,
 keeping \delr\ fixed at 0.05, with shaded areas denoting unstable 
 cases. The two regions of the instability are more clearly \bhl 
 separated than in the slice\ehl\ of fig. \ref{fig:unstable_deltar}. RI
 only occurs for very small values of \del2r\ while
 RII occurs around $\dot{m} \simeq \dot{m}_{\rm c}$, over a large
 range in \del2r, extending beyond the region of the
 graph. \label{fig:unstable_smooth}}
\end{figure}
 
Figures \ref{fig:unstable_deltar} and \ref{fig:unstable_smooth} show
the unstable regions across $[\Delta r,\dot{m}]$ and $[\Delta
r_2,\dot{m}]$. In fig. \ref{fig:unstable_deltar}, we kept $\Delta
r_2/r_{\rm in}$ fixed at 0.014 and surveyed the \bhl range\ehl\ 
$\Delta r/r_{\rm in}=[0.02,0.28]$ and $\dot{m}/\dot{m}_{\rm c}=[10^{-6},6]$. 
Figure \ref{fig:unstable_smooth} surveys the same range in \mavg\ 
and the range $\Delta r_2/r_{\rm in} = [0.004,0.14]$, keeping 
$\Delta r/r_{\rm in}$ fixed at 0.05. The figures (especially figure
\ref{fig:unstable_smooth}) show that there are two distinct regions in
which the instability is active, which we call RI and RII. 

The instability has different properties in the two regions: the period,
shape and amplitude of the outburst are all qualitatively different, 
as we demonstrate further below.

The first instability region (RI) appears in a small range of $\Delta
r_2/r_{\rm in} = [0.002,0.03]$, but over a considerable range in
\delr\ (up to 0.25) and five orders of magnitude in accretion rate:
$\dot{m}/\dot{m}_{\rm c} \simeq [10^{-6},10^{-1}]$. This instability
region was the focus of our study in DS10, where we discussed in
detail the appearance of the instability. (The phenomena summarized
below can be seen in figs. 4,5 and 6 of that paper.) The instability
is characterized by large amplitude outbursts followed by long periods
of quiescence in which the accretion rate onto the star drops to
zero. The duty cycle for the instability depends on all three
parameters (\delr,\del2r, and \mavg), decreasing significantly as the
mean accretion rate drops (and mass takes longer to accumulate and
power another cycle). The outburst profile generally takes the shape
of a relaxation oscillator, with an initial accretion peak followed by
a tail of much lower amplitude accretion. Additionally, during the
long phase of the outburst suboscillations sometimes appear in the
accretion profile. These suboscillations have a much higher frequency
than the overall burst, and appear to be the second instability (RII)
superimposed on the outburst while the mean acccretion rate onto the
star is temporarily higher. As the mean accretion rate is decreased,
the outburst becomes shorter and shorter, until finally it simply
appears as a single spike of accretion followed by a long period of
quiescence.

\begin{figure*}
      \rotatebox{90}{\resizebox{!}{120mm}{\includegraphics[width=\linewidth]{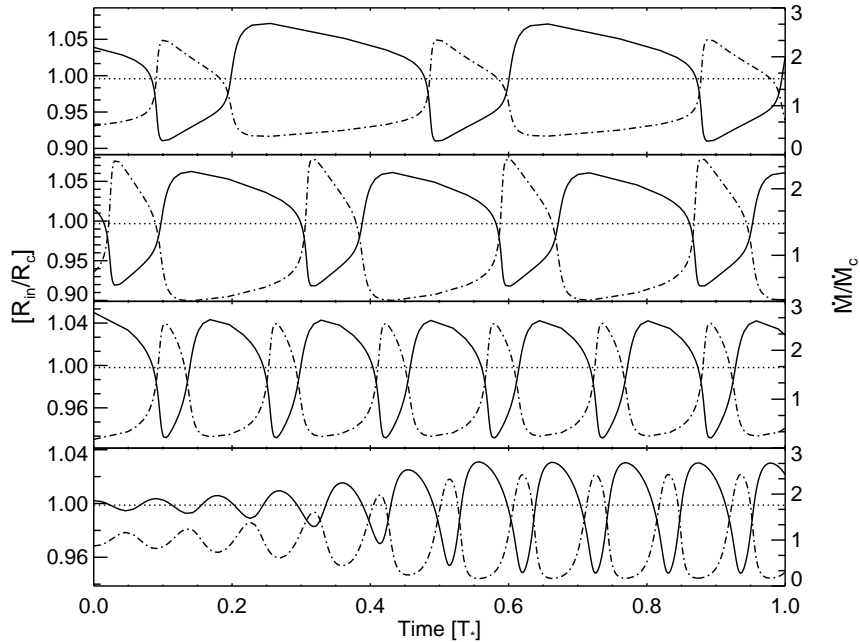}}}
      \caption{The accretion instability in RII, the second
        instability region for $\Delta r/r_{\rm in} = 0.05$ and
        $\dot{m} = 1.05~\dot{m}_{\rm c}$, with $\Delta r_2/r_{\rm in}
        = [0.028,0.05,0.1,0.15]$. In each figure, the solid line shows
        the evolution of \rin\ over a cycle, while the dot-dashed line
        shows the accretion rate scaled to units of $\dot{m}_{\rm
          c}$. The dashed line shows the inner radius corresponding to
        steady-state accretion. The instability only manifests around
        $\dot{m} \simeq \dot{m}_{\rm c}$, but occurs over a large
        range in $\Delta r_2/r_{\rm in}$, the transition length
        between accreting and non-accreting states. Unlike the RI
        instability, the accretion rate never drops to zero during the
        quiescent phase of the instability.
        \label{fig:RII_inst}}
\end{figure*}

The second instability region (RII) occurs only around
$\dot{m}/\dot{m}_{\rm c} \simeq 1$, in a relatively small range of
$\Delta r/r_{\rm in} = [0.01,0.07]$, but over a large range in $\Delta
r_2/r_{\rm in}$. (We have truncated the figure at $\Delta r_2/r_{\rm
  in}$ to make RI more prominent, but additional simulations show that
the instability continues to larger values of \del2r, up to at least
0.4). Fig \ref{fig:RII_inst} shows four sample simulations for \bhl
RII\ehl\ taken from the output of our $[\Delta r_2,\dot{m}]$ set of
simulations. The simulations all have $\dot{m} = 1.05~ \dot{m}_{\rm
  c}$ and $\Delta r/r_{\rm in} = 0.05$, with increasing \del2r\ from
bottom to top: $\Delta r_2/r_{\rm in} = [0.028,0.05,0.1,0.15]$. The
instability has a different character from the RI instability. The
quiescent phase is absent and there is always some accretion onto the
star, even during the low phase of the cycle. The shape of the
instability is also significantly different from the RI instability:
\bhl there is no additional higher frequency oscillation\ehl, and the
outburst profile is smoother. In particular, the initial spike of
accretion seen in RI is completely absent. The profile is nearly
sinusoidal for small values of \del2r\ (although the duty-cycle is
always less than 0.5). For larger values of \del2r, the outburst is
characterized by a rapid rise, followed by a decaying plateau and then
rapid decline to the low phase of the cycle.
\subsection{Period and Amplitude of Instability}
\label{sec:cycle_period}

We can use our high resolution survey of the instability parameter
space to study quantitatively how the properties of the instability
change as a function of the different parameters. This allows us to
see the differences between the RI and RII instability regions more
clearly, and better understand how the instability operates. It also
allows us to make more general predictions for the appearance of the
instability that can be tested against observations. 

To \bhl measure\ehl\ the
period of the instability we autocorrelate the output of each
simulation, and take the first peak in the autocorrelation. All our
simulations take as initial conditions the stable solution given by
(\ref{eq:sig}) that then becomes unstable. However, the instability
generally takes some time to reach a stable period and amplitude,
which varies significantly between simulations. This introduces some
error in the estimate of the period (particularly in RII, where the
instability takes longer to emerge), which is reflected in the figures
below.

Figs. \ref{fig:Period_deltar} and \ref{fig:Period_smooth} show the
instability period [top] and maximum amplitude [bottom] as a function
of accretion rate for each unstable simulation in
figs. \ref{fig:unstable_deltar} and \ref{fig:unstable_smooth}
respectively. Each individual curve represents a different value for
\delr\ or \del2r, with the colours evolving from purple (smallest
parameter value) to red (largest). In the top panels, the gaps in the
curves denote stable \bhl cases\ehl. In the bottom panels, the dashed
$y=x$ line shows the mean accretion rate for each simulation.

\begin{figure}
      \rotatebox{270}{\resizebox{!}{90mm}{\includegraphics[width=\linewidth]{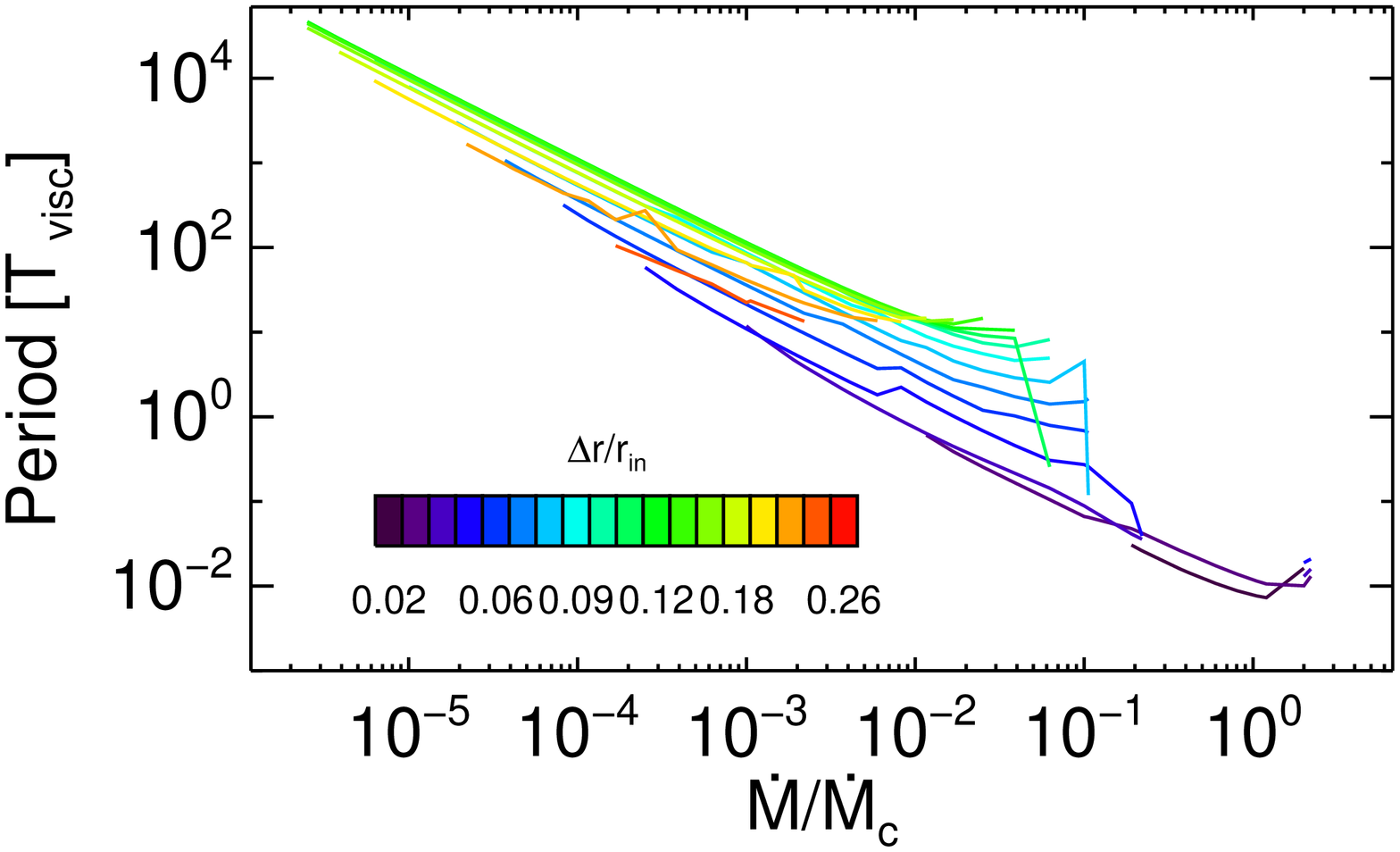}}}
      \rotatebox{270}{\resizebox{!}{90mm}{\includegraphics[width=\linewidth]{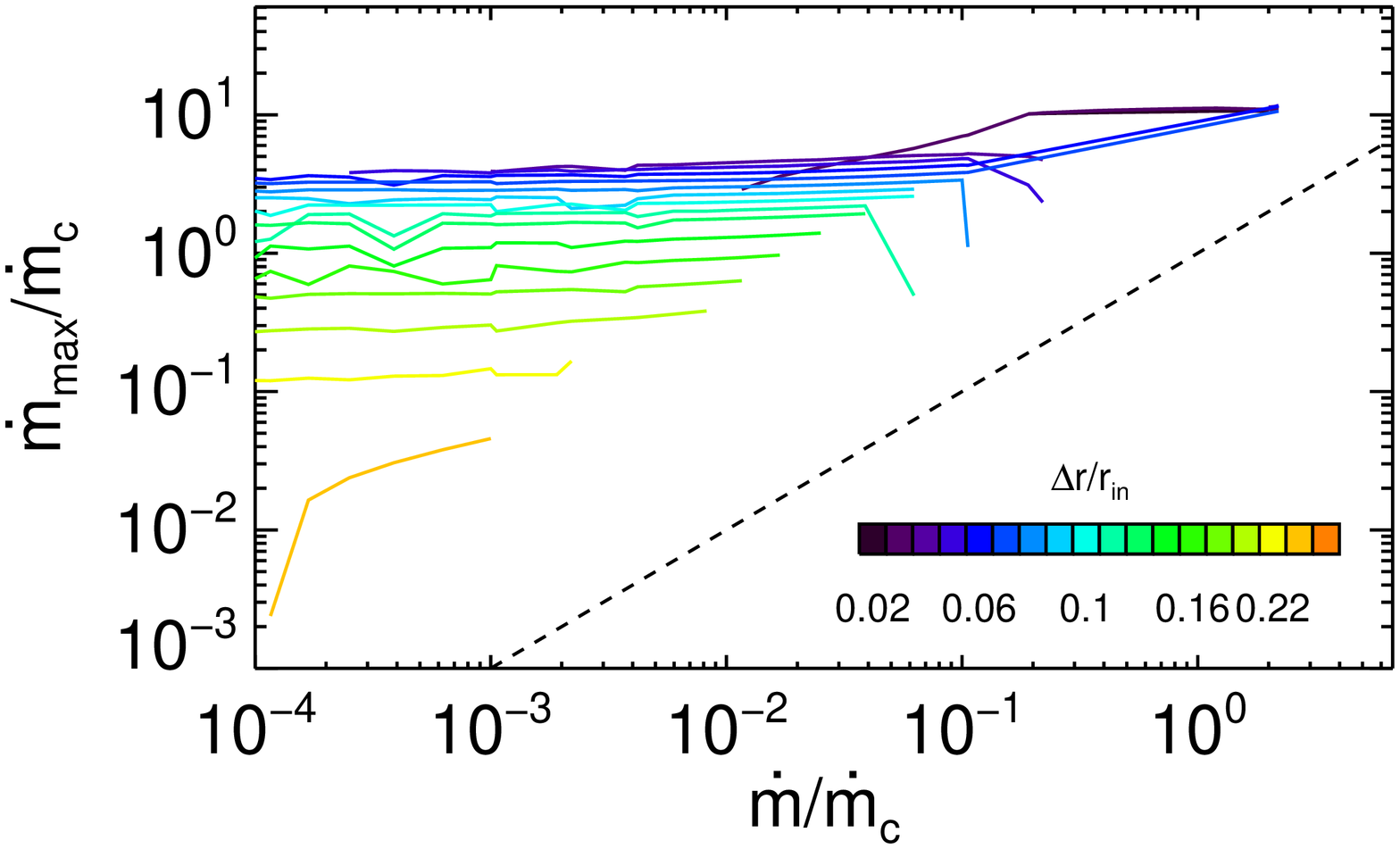}}} 
      \caption{Period [top] and amplitude [bottom] of the instability as a
        function of $\dot{m}/\dot{m}_{\rm c}$ for the unstable simulations
        shown in fig. \ref{fig:unstable_deltar}. The accretion rate spans
        the range $\dot{m}= [10^{-6},6]$ and $\Delta r/r_{\rm in} =
        [0.02,0.28]$. The individual curves show differen values of $\Delta
        r/r_{\rm in}$, increasing from black (smallest) to red (largest). In
        the top panel, the gaps in the curves indicate regions of stable
        accretion. In the bottom panel, the mean accretion rate is indicated
        by the dashed $y=x$ line.\label{fig:Period_deltar}}
\end{figure}

\begin{figure}
      \rotatebox{270}{\resizebox{!}{90mm}{\includegraphics[width=\linewidth]{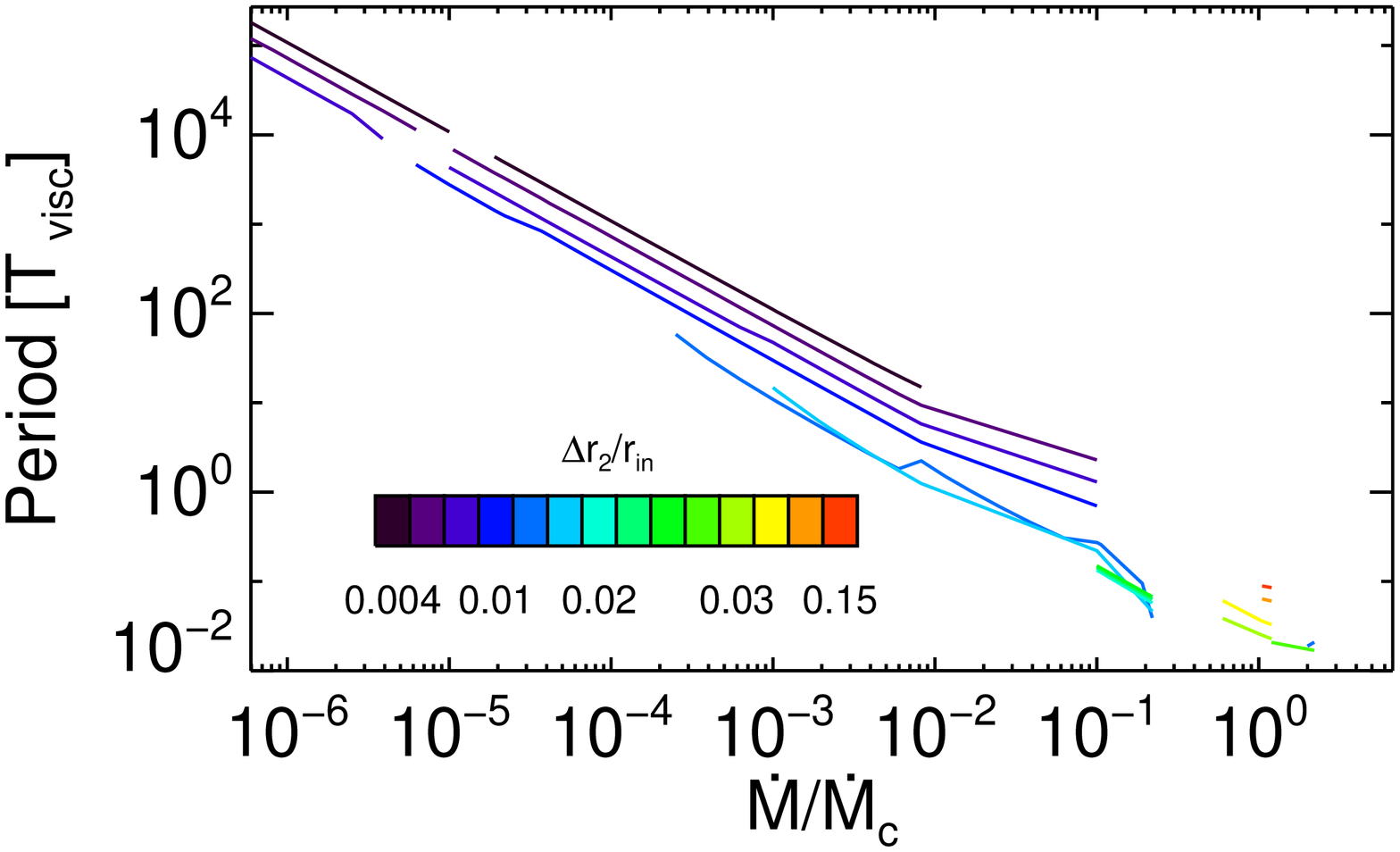}}}
      \rotatebox{270}{\resizebox{!}{90mm}{\includegraphics[width=\linewidth]{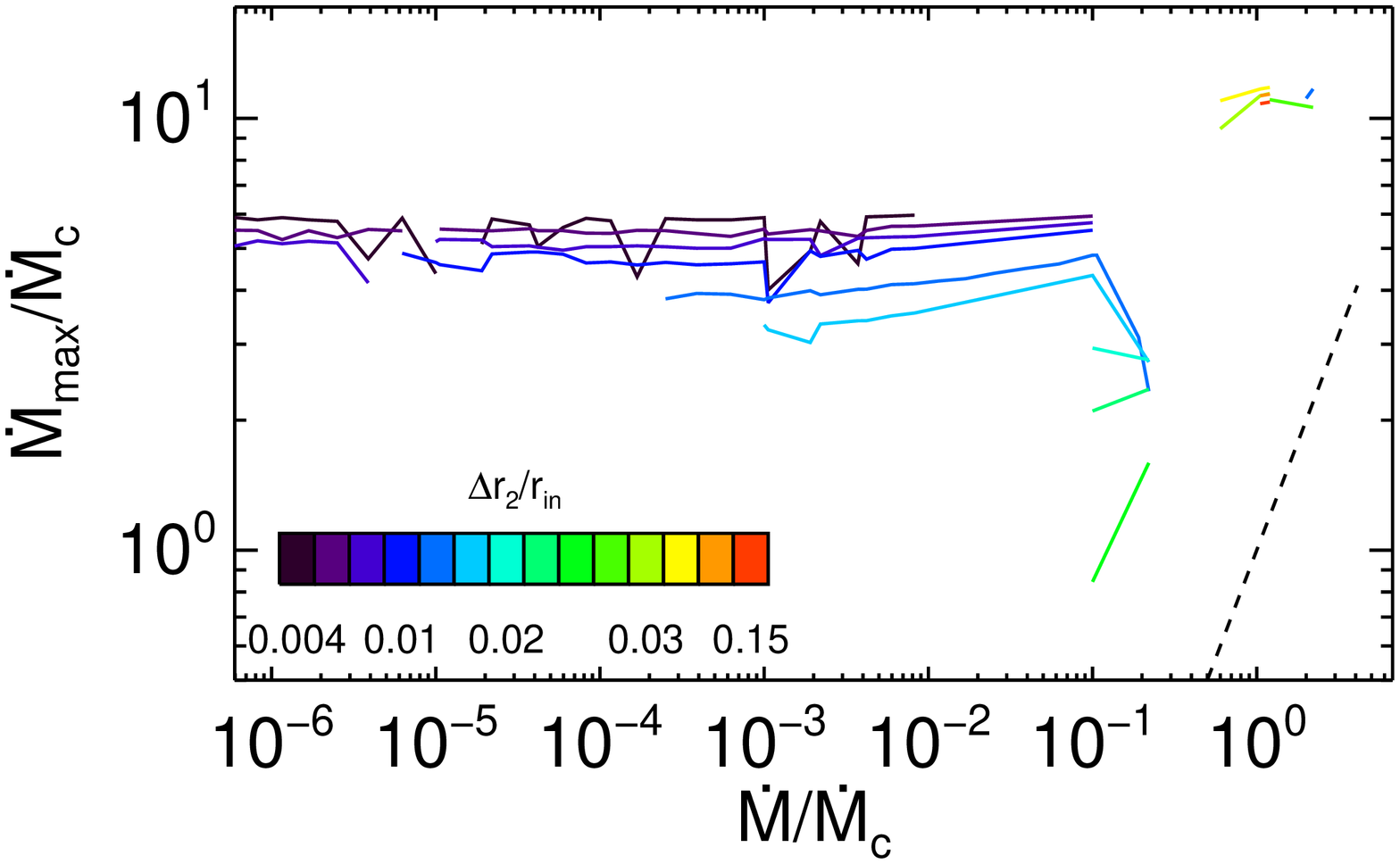}}}
      \caption{Period [top] and amplitude [bottom] of the instability
        as a function of $\dot{m}/\dot{m}_{\rm c}$ for the simulations
        shown in fig. \ref{fig:unstable_smooth}. The plot is the same
        as fig. \ref{fig:Period_deltar}, replacing \delr\ with
        \del2r. Each curve shows a different \del2r\ spanning the
        range $\Delta r_2/r_{\rm in} =
        [0.004,0.15]$.\label{fig:Period_smooth}}
\end{figure}

Some general properties of the period and amplitude of the outburst
are valid for both for RI (low \mavg;
cf. sec. \ref{sec:cycle_appearance}) and RII (high \mavg). The period
of the instability shows an approximately power-law dependence on
accretion rate, with an index independent of \delr\ and \del2r. It
also varies strongly with both \delr\ and \del2r. The period scales
directly with \delr. The maximum amplitude of the cycle also scales
roughly linearly with \delr\ and\del2r. It shows the opposite trend
for \del2r, decreasing as \del2r\ increases. On the other hand, the
amplitude of the outburst is nearly independent of the mean accretion
rate onto the star for a given \delr\ or \del2r, as is seen in the
bottom panel of figs. \ref{fig:Period_deltar} and
\ref{fig:Period_smooth}. Rather than producing a weaker outburst, at
low \mavg\ the duration of the low phase of the outburst increases as
mass accumulates in the disc.

This result demonstrates that the details of the outburst, \bhl apart 
from the period of the cycle\ehl, are largely independent of \mavg,
and depend \bhl almost\ehl\ only on \delr\ and
\del2r. For large \delr\ a larger reservoir of mass is available for
outburst, so that the duration of each burst increases. Smaller \del2r
means a more abrupt transition around \rc\ from accreting to
non-accreting states, so that for the same \mavg, a simulation with
small \del2r\ can build up more mass in quiescence than a simulation
with larger \del2r, leading to a longer outburst phase and longer
period.

Although the trends described above apply to both RI and RII,
figs. \ref{fig:Period_deltar} and \ref{fig:Period_smooth} clearly
demonstrate significant differences between the two regions. In RI the
accretion amplitude scales inversely with both \delr\ and \del2r, so
that a large \delr\ (or \del2r) corresponds to an outburst with an
amplitude up to $100\times$ smaller than for the same \mavg\ and
smaller \delr\ (\del2r). The transition between accreting and
non-accreting states becomes more gradual as \delr\ (\del2r)
increase, so it makes sense that the initial peak of the outburst will
be lower for larger \delr (\del2r). Interestingly, the cycle
amplitude scales inversely with \delr, while the period scales
directly. The opposite behaviour is true of \del2r. Instabilities
with large \delr\ thus have weaker outbursts and a longer duty cycle
(compared to an instability with smaller \delr), while those with a
large \del2r instead have weaker outbursts and a shorter period, and
less variation between outburst and quiescence.

The RII instability is presently not as well-sampled as RI, but some
general characteristics are evident. Most significantly, the amplitude
is much larger ($\sim 40\%$) than the largest amplitude RI
instability, and the period of oscillation is up to $\sim 60\%$
shorter.  As well, our present results suggest that the outburst
amplitude for RII is roughly {\em independent} of \delr, \del2r, and
\mavg, staying fixed at about $10\,\dot{m}/\dot{m}_{\rm c}$. Instead
of changing in amplitude, as $\langle\dot{m}\rangle$ decreases (or
\del2r\ increases), the period increases as well and the source spends
more time in the low phase of the cycle (as is seen in
fig. \ref{fig:RII_inst}).

In summary, the instability is very different in RI and RII. In RI,
the period of the instability is 1--1000 times longer than the viscous
accretion timescale in the inner parts of the disc, and is a strong
function of \delr, \del2r and \mavg. The amplitude and duty cycle of
the outburst are strong functions of the detailed disc-field
interaction around \rc, but appear nearly independent of \mavg. The
instability occurs over a wide range in \mavg\ and \delr, but is
confined to \bhl small\ehl\ values of \del2r, the \bhl parameter
describing the transition from accretion to `centrifugal barrier'\ehl.
RII is confined to a small range in \mavg, but extends to much larger 
values of \del2r. The cycle period is typically shorter than the viscous
timescales in the inner disc ($P_{\rm cyc} \sim 10^{-2}-1~t_{\rm c}$),
and the amplitude of the outburst is higher than RII. The amplitude of
the outburst appears to be independent of \mavg, $\Delta r$ and
$\Delta r_2$, with a \bhl typical\ehl\ value of $\sim 10\,\dot{m}_{\rm c}$.

\subsection{Interpretation of the instability regions}
\label{interp}
The instability in region I is easiest to understand; it corresponds
to the cyclic accretion predicted by \cite{1977PAZh....3..262S}, and
studied in \cite{1993ApJ...402..593S}. Mass piling up at the
centrifugal barrier eventually becomes large enough to `open the
gate', and the resulting accretion rate is large enough to push the
inner edge of the disc inside corotation for a while, until the mass
reservoir has been drained. Unlike a fixed reservoir, the mass in it
is not a fixed number, since it depends on the extent of the disc
participating in the cycle. This explains the large range in cycle
period over which it operates \citep{1993ApJ...402..593S}.

This view of the RI cycle implicitly assumes a sharp transition from
accretion to pile-up outside \rc. If the transition is not sharp and
the centrifugal barrier is `leaky', the cyclic behavior can be
avoided. Mass still piles up outside \rc, but accretion can be matched
by mass leaking through the barrier. This explains why the cycles are
limited to low values of \del2r. As in \cite{1993ApJ...402..593S}, the
cycles can start from a steady accreting state in the form a linear
instability, which sets in only if the transition is steep enough.

Region II is restricted to accretion rates around the transition to
steady accretion \bhm, and small values of \delr.\ehm\ Its restricted
range of oscillation periods -- a fraction the viscous time scale at
\rc\ -- shows that it operates on a region of finite extent, of the
order of the width of the transitions \delr, \del2r.\bhm\ In the
transition region, a small increase in accretion rate will cause \rin\
to move inwards, which will substantially decrease the torque at \rin\
(because \delr\ is small) and moves \rin\ even further inward. As \rin\
moves inward the accretion rate through \rin\ increases and an
outburst is triggered. Since even in the low phase of the cycle
material is accreted onto the star, the reservoir of matter available
for the outburst is small and only the innermost regions of the disc
are involved in the instability (of order \del2r). The timescale for
the instability is thus much shorter than the RI instability, and the
period scales with \del2r.

Since the instability occurs over a small region around \rc, it is
likely sensitive to the details of the physics of the transition
region. As long as we may assume the transition to be monotonic,
however, it's shape is rather constrained by its values far from
corotation (0 and 1 respectively), and its slope at corotation,
(controlled by the width parameter, $\Delta r_2$). While the results
above show that the existence and properties of the cycle depend
strongly on $\Delta r_2$, one might wonder to what extent the results
also depend on remaining degrees of freedom in the choice of
transition function. To test this, we re-ran a series of simulations
using the error function rather than the $\tanh-$ function. Since the
width of the function is a matter of definition in both cases, we use
slope of the function at corotation as a well-defined measure of width
for the case-by case comparison. The range of unstable values in the
parameter space, the period, shape, and amplitude of the cycle were
then found to be very similar as when using a $\tanh$ profile.\ehm

\begin{figure}
      \rotatebox{270}{\resizebox{!}{90mm}{\includegraphics[width=\linewidth]{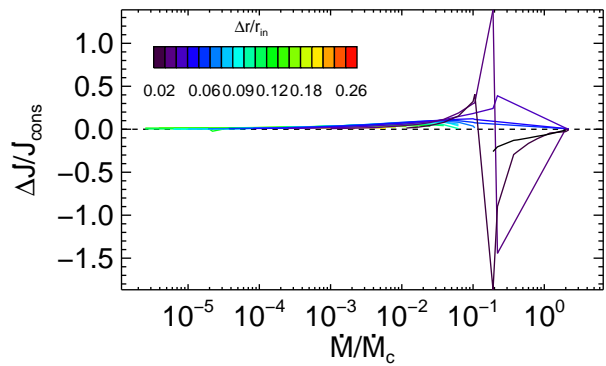}}}
\caption{\bhl Effect of cycles on the average torque, as a function 
of accretion rate. Figure shows the difference between the torque in 
the unstable initial state before the cycle develops, and the value measured 
in the fully developed limit cycle\ehl. The different curves represent
 different \delr, increasing from purple ($\Delta r = 0.05$) to 
red ($\Delta r = 0.22$). $\Delta r_2 = 0.014$ is held constant.
 \bhl Cases that do not become cyclic lie on the red line along 
 the x-axis\ehl. 
 \label{fig:dotm_deltar}}
\end{figure}

\begin{figure}
      \rotatebox{270}{\resizebox{!}{90mm}{\includegraphics[width=\linewidth]{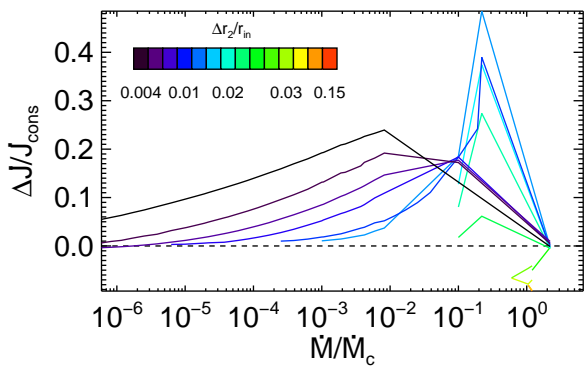}}}
\caption{\bhl Difference between the torque in the unstable initial state 
before the cycle develops, and the value measured 
in the fully developed limit cycle\ehl. The different curves represent
 different \del2r, increasing from purple ($\Delta r_2 = 0.004$) to
 red ($\Delta r = 0.028$). $\Delta r = 0.05$ is held
 constant. Solutions that lie along the x-axis do not show cyclic
 variations.\label{fig:dotm_smooth}}
\end{figure}

\begin{figure}
  \rotatebox{270}{\resizebox{!}{90mm}{\includegraphics[width=\linewidth]{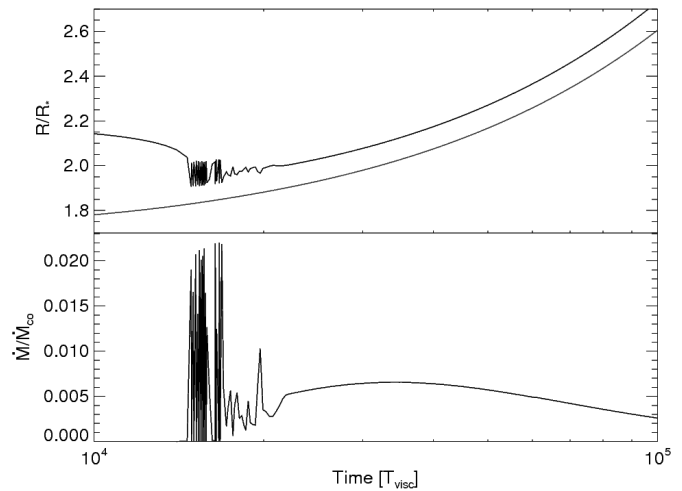}}}
  \caption{Unstable disc solution for $[\Delta r,\Delta r_2] =
    [0.2,0.02]r_{\rm in}$. At $t=0$, $r_{\rm in}/r_{\rm c} = 1.3$, so
    that $\dot{m} \simeq 0$. Initially \rin\ remains constant, before
    decreasing sharply as \rc\ increases. This prompts the onset of
    the accretion instability, which continues even while the star
    spins down. At later times, the amplitude and frequency of the
    instability decrease and the disc reverts back to accreting
    steadily. Top: Evolution in \rin\ (black curve) and \rc\ (red
    curve), before, during, and after the instability appears. Bottom:
    Corresponding accretion rate onto the star. The accretion rate is
    given in terms of $\dot{m}_{\rm c}$ corresponding to \rc\ at
    $(t=0)$.\label{fig:transient}}
\end{figure}

\subsection{The effect of cycles on angular momentum exchange}
\label{sec:cycle_jdot}
The accretion instability can have a significant effect on the spin
change with the star. When accretion is stable, the rate of angular
momentum exchange between the star and the disc is given by
(\ref{eq:Jdot}), and illustrated in fig. \ref{fig:nu_mdot}. When
accretion proceeds via cycles, the amplitude of the outburst is
generally much higher than the mean \mavg, and the disc spends a
significant amount of time in the low phase of the cycle, in which
little or no accretion takes place (sec. \ref{sec:cycle_period}). This
will affect the net angular momentum exchange between the star and the
disc: since the low phase of the instability almost always lasts
longer than the outburst phase, the star will spin down faster than if
accretion proceeded steadily.

Using the results of the simulations presented in
sec. \ref{sec:cycle_appearance} and \ref{sec:cycle_period}, we can
compare the torque exerted on the star for stable and unstable
accretion:
\begin{equation}
\frac{\Delta \dot{J}}{\dot{J}_{\rm cons}} \equiv \frac{\langle\dot{J}_{\rm
 sim}\rangle-\dot{J}_{\rm cons}}{\dot{J}_{\rm cons}}.
\end{equation} 
Here $\dot{J}_{\rm cons}$ is the torque exerted on the star \bhl
in the initial state\ehl\ (calculated using \ref{eq:Jdot}), and
$\langle\dot{J}_{\rm sim}\rangle$ is the average torque exerted over
one full cycle \bhl once a stable limit cycle is reached\ehl.

Figs.~\ref{fig:dotm_deltar} and \ref{fig:dotm_smooth} show \Djdot\
as a function of accretion rate for the simulations in
figs. \ref{fig:unstable_deltar} (varying \delr) and
\ref{fig:unstable_smooth} (varying \del2r), respectively. The
individual curves show different \delr\ (or \del2r), with the colours
from purple to red indicating increasing values of the parameter. Lines
along the x-axis are stable solutions.

The results are easily separated into the RI (low \mavg) and RII (high
\mavg) instability regions. In RI, both $\dot{J}_{\rm cons}$ and
$\dot{J}_{\rm sim}$ are negative (cf. \ref{fig:nu_mdot}) over
the entire unstable region, and the star spins down. As expected, the
presence of the instability increases the efficiency of the spin-down
torque considerably: by up to $50\%$ for $[\Delta r/r_{\rm in},\Delta
r_2/r_{\rm in}] = [0.05,0.015]$. For a given value of \delr\ or \del2r,
\Djdot\ is largest at the highest \mavg\ and decreases for smaller
\mavg. When \mavg\ stays fixed, \Djdot\ is largest for the smallest values
of \delr\ and \del2r (where the outburst amplitude is largest,
cf. sec. \ref{sec:cycle_period}). This behaviour is readily
understandable: the added angular momentum from accretion is most
relevant at high \mavg, and the rate at which this material is
accreted (i.e. the duty cycle of the outburst) will determine how much
the star will spin up or down. The larger the amplitude of the
outburst, the faster the reservoir of mass will be accreted, and so
the shorter the duty cycle.

 In RII the situation is quite different. In the parameter range for
 \delr\ and \del2r\ where the solution is unstable,
 $\langle\dot{J}_{\rm sim}\rangle$ and $\dot{J}_{\rm cons} > 0$ when
 $\dot{m} = \dot{m}_{\rm c}$ (fig. \ref{fig:nu_mdot}). The presence of
 the instability decreases the rate of spin-up by a maximum by up to
 about 10\%.
\subsection{Transient instability cycles}
\label{sec:cycle_transient}

In DS11 we studied how discs in our model evolved as the spin of the
star changed. To do this we introduced the star's moment of inertia,
$I_*$ as an additional parameter of the problem, which introduced a
new characteristic timescale for the problem, the spindown timescale
$T_{\rm SD}$:
\begin{equation}
\label{eq:tsd}
T_{\rm SD} \equiv P_*/{\dot P}_* \sim \frac{I_*\Omega_*r_{\rm in}^4}{\eta\mu^2\Delta r}.
\end{equation}
We found that the ratio of $T_{\rm visc}$, the viscous accretion
timescale at \rc, and $T_{\rm SD}$ plays an important role in
determining whether a disc will become trapped with \rin\ close to
\rc\ when \mavg, the mean accretion rate through the disc drops to
zero. 

The spin change of the star can also determine whether or not the
accretion instability occurs. For a given set of $[\Delta r,\Delta
r_2]$ the instability frequently appears over a narrow range in
$\dot{m}/\dot{m}_{\rm c}$, so that changing \rc\ (and hence
$\dot{m}_{\rm c}$) will cause the instability to appear or disappear. 
This effect is difficult to observe in our
simulations, since the timescale of the instability ($\sim
10^{-2}-10^{3}~T_{\rm visc}$) is in general much shorter than the
spin-down timescale of the star ($T_{\rm SD} \sim 10^{3}-10^{17}~
T_{\rm visc}$; DS11), so once the instability occurs the simulation typically
proceeds on the instability timescale. Despite this, we do find some solutions that show
transient oscillations (typically those with the long periods, $P \sim
10^{3} T_{\rm visc}$) for a constant $\dot{m}$, which appear and
disappear as \rc\ changes. 

\begin{figure*}
  \rotatebox{90}{\resizebox{!}{120mm}{\includegraphics[width=\linewidth]{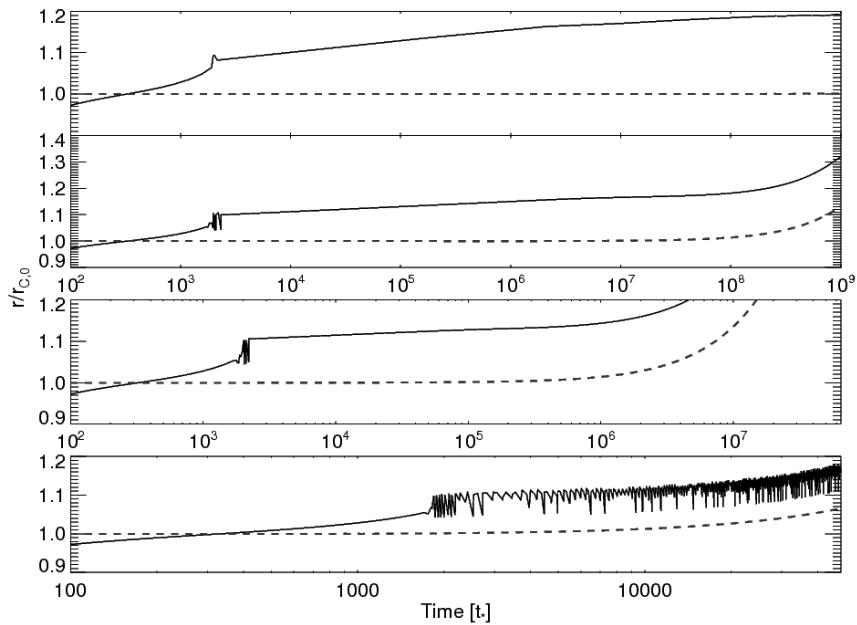}}}
  \caption{Evolution from an accreting to a non-accreting disc, for
    increasing (top to bottom) ratios of $T_{\rm visc}/T_{\rm SD}$ for
    stable discs ($\Delta r/r_{\rm in} = 0.1$, $\Delta r_2/r_{\rm in} =
    0.04$). From top to bottom, the ratio $T_{\rm visc}/T_{\rm SD}$ is
    $2.5\times [10^{-9},10^{-7},10^{-5},10^{-4}]$. The instability
    appears in all four panels, for varying lengths of time. As the
    ratio between the two timescales decreases, the disc is not able to
    move outwards as quickly before the star begins to spin down, so
    that \rin\ will always remain close to \rc\ and the instability
    persists for longer. \label{fig:transient_visc} }
\end{figure*}

This transient instability is illustrated in figs. \ref{fig:transient} and
\ref{fig:transient_visc}. For these simulations we have adopted the same
scaling parameters as we used in DS11. We take $r_*$, the star's
radius, as our characteristic length scale, and define $t_* \equiv
\nu/r^2_*$, the nominal viscous timescale at the star's radius, as the
characteristic timescale. We set $T_{\rm SD} \equiv
380\,T_{\rm visc}$ (where $T_{\rm visc}$ is the viscous
timescale for $r_{\rm c}(t=0)$), which is appropriate for a
protostellar system. We set $r_{\rm c}(t=0) = 1.8 r_*$.

In fig. \ref{fig:transient} we show the evolution of an initially
dead disc ($\dot{m} = 0$) with $[\Delta r/r_{\rm in},\Delta r_2/r_{\rm
  in}] = [0.2,0.02]$. The disc is initially in the steady-state
solution given by (\ref{eq:sig}) with $r_{\rm in} = 1.3 r_{\rm
  c}$. The top panel shows the evolution of $r_{\rm in}$ (black) and
$r_{\rm c}$ (red) in time, while the bottom panel shows the accretion
rate onto the star, scaled to $m_{\rm c}$ at $t = 0$. At early times
[not shown] there is no accretion onto the star and \rin\ remains
fixed while \rc\ moves steadily outwards. Once \rc\ moves close enough
to \rin\ (at $\sim10^{4}t_*$) to begin accreting matter, gradients in
the disc's surface density cause \rin\ to move quickly closer to
\rc. At $r_{\rm in}/r_{\rm c} \sim 1.1$, however, the disc suddenly
becomes unstable and accretion proceeds via a series of large
amplitude bursts, which increase the maximum accretion rate onto the
star enormously. \rc\ continues to move outwards as the source
oscillates, which causes $\dot{m}_{\rm c}(r_{\rm c})$ to decrease as
well. Eventually $\dot{m}/\dot{m}_{\rm c}$ moves out of the unstable
range, and the disc settles back into a steadily accreting solution
with a gradually decreasing accretion rate. This result emphasizes the
fact the instability can depend sensitively on $\dot{m}/\dot{m}_{\rm
  c}$, so that as \rc\ evolves, cycles can appear and then disappear.

For constant \delr\ and \del2r, the appearance of cycles depends on
\mavg\ and $\dot{m}_{\rm c}$. As we showed in DS11, when the mean
accretion rate through the disc drops so low that the disc becomes
trapped, the accretion rate at \rin\ depends strongly on $T_{\rm
  visc}/T_{\rm SD}$. This ratio also plays an important role in
determining the presence and duration of unstable accretion phases.

We illustrate this in fig. \ref{fig:transient_visc}, which shows a
disc evolving from an accreting disc to a trapped disc for different
$T_{\rm visc}/T_{\rm SD}$, adopting parameters appropriate for a
neutron star. We take the field strength $10^{12}$ G, initial spin
period $5{\rm s}$, and initial inner edge radius $0.95r_{\rm c}$. The
disc is thus initially in an accreting state. The outer boundary is
located at $100~r_{\rm in}$. The initial spindown timescale of the star
is $T_{\rm SD} = 10^5$ years.  We adopt the same system parameters as
for fig. \ref{fig:transient}, with $[\Delta r/r_{\rm in},\Delta
r_2/r_{\rm in}] = [0.2,0.02]$. At $t=0$ we set $\dot{m} = 0$ at the
outer boundary of the disc, so that the accretion rate through the
disc gradually drops.

The figure shows the evolution of \rc\ (dashed curve) and \rin\ (solid
curve) as a function of time for increasing $T_{\rm visc}$, with
$T_{\rm visc}/T_{\rm SD} = 2.5\times
[10^{-9},10^{-7},10^{-5},10^{-4}$] from top to bottom. In all four
simulations, the disc initially move outwards as \mavg\ decreases
to about $1.05\,r_{\rm c}$, where the accretion rate triggers the onset
of the instability. 

The appearance of the disc changes dramatically with changing
viscosity. Discs with a high viscosity are able to diffuse outwards
quickly as \mavg\ declines, so that \rin\ in the top panel (with the
highest viscosity) moves far from \rc\ before the star can begin to
spin down, and the disc never becomes trapped close to \rc. The
instability (which only occurs between $\dot{m}/\dot{m}_{\rm c} \simeq
[10^{-3},0.1]$) will only appear briefly before \rin\ moves too far
from \rc\ and the accretion rate drops. In the middle two panels the
instability also appears briefly, but the star does begin to spin down
quickly enough that \rc\ begins to move outwards at the same rate as
\rin\ and the disc remains trapped. In the bottom panel (with the
lowest viscosity), the disc diffuses outwards slowly, and the star
spins down more rapidly. Once the instability is triggered, \rin\
remains close enough to \rc\ that the instability persists as the star
continues to spin down.

\section{Relevance for Astrophysical Sources}
\label{sec:sources}
\subsection{Trapped discs and the disc instability in EXors}
\label{sec:ttauri}


We have previously suggested that the disc instability discussed in
this paper is operating in a class of T~Tauri stars known as
`EXors'. EXors, like their prototype, EX Lupi are characterized by
repeated large outbursts: changes by up to four magnitudes in
luminosity lasting several months, with a characteristic total period
of several years \citep{2007AJ....133.2679H,
 2008AJ....135..637H}. Observations of these sources in both outburst
and quiescence can produce valuable constraints on both the disc
instability and the trapped disc itself.

\begin{figure}
      \rotatebox{90}{\resizebox{!}{90mm}{\includegraphics[width=\linewidth]{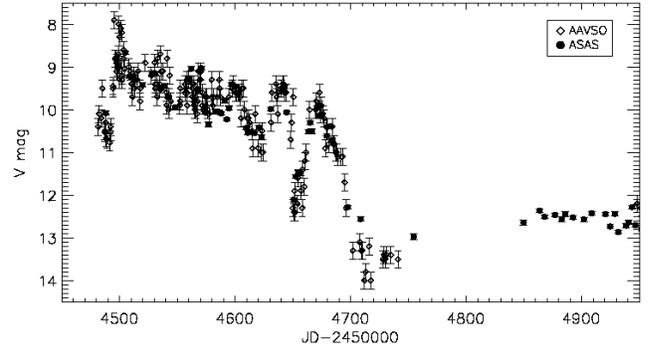}}}
\caption{The 2008 outburst of EX Lupi. From January to August, the
  source was up to 4 magnitudes larger than its quiescent value. {\em
    Figure courtesy A. Juhasz}.\label{fig:EXLup}}
\end{figure}

Figure \ref{fig:EXLup} shows the $m_{\rm V}$ brightness of EX Lupi
over 500 days, taken from the AAVSO and ASAS archives (from Attila
Juhasz, {\em private communication}). In January 2008, EX Lupi
underwent a large outburst, increasing by 3 magnitudes, from $m_{\rm
  V} = 11$ to $m_{\rm V} = 7.9$ \citep{2010ApJ...719L..50A} in about
20 days. In July 2008 EX Lupi was at $m_{\rm V} =12.4$
\citep{2009A&A...507..881S}, so the total change in luminosity was more
than 4 magnitudes. EX Lupi remained in outburst for eight months,
decaying from maximum to about $m_{\rm V} = 10$ before abruptly
dropping to $m_{\rm V} = 14$ over about 40 days. The data show
evidence of an additional higher-frequency modulation with a period of
about 30 days and amplitude of 1-2 magnitudes.

The outburst profile looks similar to some of the profiles seen in our
simulations, with a rapid rise and a gradual decreasing plateau phase,
followed by a rapid decay as the source returns to quiescence. The
higher frequency modulation is also suggestive of the higher frequency
oscillations seen in the RI instability discussed above and in DS10,
although its short timescale $\sim 30$ days, or $\sim 4-10\times$
longer than $P_*$ \citep{2009A&A...507..881S} might also indicate
direct modulation between the disc and magnetic field
(e.g. \citealt{1997ApJ...489..199G,2008ApJ...673L.171R}). 

Spectral observations taken both during the outburst and quiescence
allow additional constraints on the disc
model. \cite{2010ApJ...719L..50A} analysed a series of spectra of EX
Lupi from the 2008 outburst to constrain the parameters of the
disc. They calculated an accretion rate of $2\times10^{-7} M_\odot
{\rm yr}^{-1}$ in outburst, and modelled the 2.2935$\mu$m CO overtone
emission to suggest an inner gap of about $10~r_*$ in the disc. They
also used previously published results to calculate $6\times10^{-9}
M_\odot {\rm yr}^{-1}$ for the quiescent
emission. \cite{2009A&A...507..881S} analysed a series of spectra for
EX Lupi in quiescence (with $m_{\rm V} \sim 12$), and did a detailed fit for
the spectrum. They estimated an accretion rate of $4\times10^{-10}
M_\odot {\rm yr}^{-1}$, and fit the spectrum with a dust-dominated disc
truncated at about 0.2 AU. These observations thus suggest that the
accretion rate increased by $\sim 30-500\times$ during the outburst,
and that close to the star, gas is either absent or optically thin.

To see whether these observations are compatible with our model of a
trapped disc, we adopt the best-fit parameters of
\cite{2009A&A...507..881S}, with $M_*=0.6M_\odot$, $r_* = 1.6R_\odot$,
and $P_*=6.3d$ (based on $v\sin i = 4.4 km/s$ and an inclination of 20
degrees). The magnetic field is unknown, so we assume a strong dipolar
field with $B_* = 1500G$, which is comparable to field strengths
observed in T~Tauri stars such as BP Tau
\citep{2008MNRAS.386.1234D}. The rest of the parameters we adopt are
the same as for our representative model
(sec. \ref{sec:standard}). This puts the co-rotation radius at
$7.6~r_*$, and sets $r_{\rm in} \simeq r_{\rm c}$ for the quiescent
disc. The viscous timescale at \rc\ is thus $\sim 30~\rm{yrs}$. If we
assume that the period of the instability is $\sim 60~\rm{yrs}$ (the
time between the 2008 outburst and the similar amplitude outburst in
1955), or $\sim T_{\rm visc}$. The instability would then fall in the
RII region.

Additional information could come from directly detecting the
accretion disc itself in quiescence. A trapped disc will have both a
higher surface density and temperature than a standard accreting disc
with the same $r_{\rm in}$. For EX Lupi in quiescence (using the
parameters given above), our model predicts a surface temperature from
internal viscous dissipation of 530-650K (depending on \mavg),
compared with 290-570K for a Shakura-Sunyaev disc. This modest
temperature increase is likely insignificant compared to heating from
the central star, which will thus still determine the gas behaviour at
these low accretion rates (e.g. \citealt{2004ApJ...617..406M}). The
surface density in the disc will also increase modestly, from a
maximum of between 3-40 g\,cm$^{-2}$ ($\rho \simeq
4\times10^{-11}-6\times 10^{-10}$ g\,cm$^{-3}$ to 40-80 g\,cm$^{-2}$
($\rho \simeq 4\times10^{-10}-10^{-9}$ g\,cm$^{-3}$). At these low
densities the disc is probably optically thin
(e.g. \citealt{2010ARA&A..48..205D}), so that it is likely not
possible to directly distinguish between a trapped disc and normal
accreting one except through detailed modelling of spectral features
in the disc.

Unlike the accretion instability (which only occurs under
certain conditions in the disc-field interaction), the build-up of
mass just outside the co-rotation radius should be generic in
stars with strong magnetic fields and relatively low accretion
rates. The best opportunity to detect a trapped disc directly would
come from a star with a strong dipolar magnetic field, high spin rate
and low accretion rate, so that the disc would be optically thick
rather than thin (as would be expected for the given accretion rate),
and could be directly detected through modelling of its spectral
features. 

\subsection{NS transients w/ weak recurrent outbursts}
\label{sec:transient}

In DS11 we suggested that the large ratio between the spin-down and
viscous timescales in accreting millisecond pulsars makes it unlikely
that the disc will become trapped, since as \mavg\ drops \rin\ moves
outwards much faster than \rc. This agrees with the typical outbursts
seen in transient X-ray pulsars, which show outbursts of variable
duration followed by long periods of quiescence and are believed to be
triggered by ionization instabilities in the disc
(e.g. \citealt{2001NewAR..45..449L}).

However at least two accreting X-ray pulsars, NGC 6440 X-2
\citep{2010ApJ...714..894H} and IGR J00291+5934
\citep{2010arXiv1006.1908H} have been observed to undergo weak,
recurrent outbursts on much shorter timescales (around 30 days) than
would be predicted from an ionization-instability model
\citep{2001NewAR..45..449L}. The weakness of the outburst has been
interpreted to mean that the disc is never completely ionized during
the outburst \cite{2010arXiv1006.1908H}, which suggests the ratio of
$T_{\rm SD}/T_{\rm visc}$ becomes small enough to trap the disc as
when the accretion rate drops. This large reservoir of matter would
then be present when the accretion rate again increases, and so could
allow more frequent outbursts. These sources thus offer an excellent
opportunity for comparison with our model of a trapped disc.

IGR J00291+5934 is a 599Hz X-ray pulsar in a 147 minute orbit, which
historically has shown standard single outbursts (in 1998, 2001, and
2004) followed by long periods of quiescence. However, in 2008 it
underwent a substantially different outburst, with two weaker
outbursts (respectively lasting nine and 13 days) separated by 27 days
of quiescence \citep{2010arXiv1006.1908H}. The authors estimated that
the total mass flux of weaker outbursts together was approximately
equal to the single 2004 outburst. Similar outbursts have been seen in
NGC 6440 X-2, a recently discovered X-ray binary with a frequency of 442Hz in a
57 minute orbit \citep{2010ApJ...712L..58A}. Since discovery on July
28, 2009, five distinct outbursts have been observed, each typically
lasting around four days at a luminosity of around $L_X \sim 10^{35} {\rm
 erg~s^{-1}}$, with a minimum recurrence time of around 31 days.

The short recurrence timescales suggest that a considerable amount of
mass is stored in the disc when the accretion rate drops, which can
then fuel another outburst when \mavg\ increases again. In both
sources the quiescent X-ray luminosity is at least 4 orders of
magnitude lower than the outburst luminosity, indicating a large
change in accretion rate (although the X-ray luminosity may not be a
good indicator of bolometric luminosity at such low accretion rates,
so there is some uncertainty in the change in \mavg).

During two outbursts, the power spectrum of NGC 6440 X-2 was dominated
by a strong low-frequency QPO ($\sim 1$ Hz)
\citep{2010ATel.2672....1P,2010arXiv1006.1908H}. A similar QPO was
previsouly detected in the tail an outburst in X-ray pulsar SAX
J1808.4-3658 \citep{2009ApJ...707.1296P} which was attributed to the
disc instability presented in DS10 and \cite{1993ApJ...402..593S}. A
much weaker QPO at $0.5~\rm{Hz}$ was observed in the outbursts IGR
J00291+5934 and might also be attributed to the disc instability
\citep{2010arXiv1006.1908H}.

The fact that all of these QPOs were observed at a similar frequency
presents a challenge to our interpretation. The outbursts seen in SAX
J1808.8-3658 follow the more conventional pattern of a large outburst
followed by a long quiescence, which would suggest that a dead disc
does not form. However, the similarity of the QPO frequency (which is
roughly correlated with the viscous timescale near \rin\ in our model)
suggests that the viscous timescale is similar in all these
sources. This contrasts with our suggestion that the viscosity in
these weak outbursts is substantially smaller than in larger
outbursts. The number of uncertain parameters in our model (in
particular this assumes that \del2r\ is similar in all sources) means
that the picture we have presented could still be valid, but then the
similarity in QPO frequency in both (or all three) of these sources is
probably coincidental. On the other hand, a strong low-frequency QPO
has currently been observed in only two sources, while as
\cite{2010ApJ...712L..58A} have noted, weakly outbursting sources like
NGC 6440 X-2 will generally be missed in current surveys.

In our present model, the disc-field interaction is not responsible
for directly regulating the accretion rate onto the star to control
the outbursts, as was suggested by \cite{2010arXiv1006.1908H}. The
field can temporarily halt accretion onto the star, but only on the
much shorter timescale on which the 1Hz QPO is seen. Thus the
mechanism that directly triggers an accretion outburst remains
undetermined.

\subsection{Persistent X-ray pulsars}
\label{sec:persistent}

A small subset of accreting pulsars with low-mass companions have
sufficiently persistent X-ray emission to measure the spin change
directly and study the connection between spin change and luminosity.
Three well-studied persistent sources (Her X-1, 4U 1626-67, and GX
1+4) all show episodes of spin-up and spin-down, although the pattern
of spin-up and spin-down is different in all three
\citep{1997ApJS..113..367B}. A fourth, (GRO J1744-28) shows clear
spin-up when in outburst, allowing the relationship between
$\dot{\nu}$ and \mavg\ to be constrained.

The binary properties of all these sources varies considerably, both
in their companions and orbital periods. However, it is still possible
to make some general observations that apply to all sources. First,
there is a clear correlation between luminosity and spin change, with
higher luminosity corresponding to spin up, and vice versa. Second,
when the star is spinning up, the spin-up rate is substantially lower
than would be predicted from the X-ray luminosity. Third, where the
relationship between $\dot{\nu}$ and \mavg\ can be measured,
$\dot{\nu} \propto \dot{m}^\gamma$, where $\gamma > 6/7$, i.e. the
scaling expected for $r_{\rm in} \propto \dot{m}^{-2/7}$. Finally, the
spin-up torque is often comparable (within a factor 2) to the
spin-down torque, with only a change in sign.

The first observation is universally expected in magnetospheric
accretion, since in order to spin up the star the matter must be able
to overcome the centrifugal barrier imposed at \rc. The second point
is also readily understandable in our model, which assumes that there
is a transition region for $r_{\rm in} \sim r_{\rm c}$ in which there
is both accretion and spin-down torque. This behaviour is also seen in
MHD simulations of the disc-star interaction
(e.g. \citealt{2004ApJ...616L.151R}). For high accretion rates, our
model predicts $\dot{\nu} \propto \dot{m}^{9/10}$, which is close to
the dependence measured in some X-ray pulsars in outburst. For
example, the 1995/1996 outburst of GRO J1744-28 (a disc-fed binary),
was measured to follow $\dot{\nu} \propto \dot{m}^{0.96}$
\citep{1997ApJS..113..367B}. This is closer to observation than the
standard model, which predicts an index of $6/7$. 

The final point is most puzzling in a model in which $r_{\rm in}
\propto \dot{m}^{-2/7}$, because it requires that the change in
accretion rate is finely tuned to preserve this symmetry. It is more
plausible for our model, however, which shows a steep decrease in
$\dot{\nu}$ around $\dot{m}/\dot{m}_{\rm c} = 1$, and then a
flattening around the minimum in $\dot{\nu}$. Thus, a small decrease
in \mavg\ will cause an initial steep transition to spin-down before
settling around the minimum in $\dot{\nu}$. There is no comparable
turnover at high \mavg, but in order to switch between spin-up and
spin-down, the star must have an average accretion rate such that
$r_{\rm in}\sim r_{\rm c}$, while in order to be a persistent source,
the accretion rate at large distances cannot be very variable. If we
assume that the average accretion rate has been steady enough on long
timescales to affect the star's rotation rate, then there will be a
natural spin-down torque, set by the minimum in \ref{fig:nu_mdot}, and
we would expect that on average the accretion rate would fluctuate
around the level that keeps \rin\ close to \rc. This still does not
account for the observations (for example of 4U 1626-27) in which the
magnitude of the torque is very nearly symmetric.

Our present model is too simple to make quantitative comparisons to
data. For one thing, the relation between $\dot{\nu}$ and \mavg\
depends on the least certain parts of our model: both the transition
widths \delr\ and \del2r\ and the connecting functions $y_m$
and $y_\Sigma$. There is also much more variation in the luminosity at
which the system moves from spin-up to spin-down than is generically
predicted in any simple magnetospheric accretion model. This
variability could indicate fluctuations in $\langle\eta\rangle$, and
\delr, which would both affect the magnitude of spin-down
torque. \cite{2000MNRAS.317..273A} demonstrated that the value of
$B_\phi$ changes if field lines are able to diffuse radially through
the disc. The diffusion of field lines is expected to happen on
viscous timescales \citep{2009A&A...507...19F}, so that the details of
the disc-field interaction could in fact rely on the complicated
interplay between the accretion rate and the field strength and
geometry. Such questions require detailed MHD modelling which are
beyond the scope of the present work.

\section{Discussion}
\label{sec:discussion}

At low accretion rates, the strong magnetic field of a star can
strongly alter the structure of an accretion disc, preventing it from
either accreting or expelling material outward, and preventing the
inner edge of the disc from moving considerably away from \rc. As we
have discussed in this paper, this disc state can lead to
substantially different behaviour from what is conventionally assumed,
and could explain a variety of observed phenomena in magnetically
accreting stars. However, in order to construct our model we used a
parameterized description for the interaction between the magnetic
field, introducing the free parameters \delr, and \del2r. Here we
discuss what sets the value of these parameters, and whether these
parameters could vary in time, which could account for some of the
additional behaviour not described by our current model.

The first parameter, \delr\, is the width of the direct interaction
between the field and the disc in the disc's innermost region. We have
assumed in all cases that $\Delta r/r_{\rm in} < 1$, so that the
interaction region (where the field lines remain closed 
\bhl long enough to exert a significant stress on the disc\ehl) remains 
fixed. MHD simulations suggest that this region will fluctuate in time 
as field lines open and reconnect, but we have assumed that this 
will occur on timescales of $\sim P_*$, which are much shorter than 
the viscous timescale of the instability ($T_{\rm visc} > 10^3 P_*$). 
Numerical MHD simulations do sometimes show variability
on longer timescales (e.g. \cite{2009arXiv0907.3394R}), which could
lead to changes in \delr\ between individual outbursts. Given that
even small changes in \delr\ can significantly change the appearance
and amplitude of the outburst (seen in figure~\ref{fig:Period_deltar}),
variability in the disc-field interaction could produce variability in
the appearance of the instability, especially if the instability
occurs in RII.

A second possible source of variability could come from the radial
evolution of the magnetic field itself. This was studied by
\citep{2000MNRAS.317..273A}, who found that radial field-line
diffusion through the disc can reduce the effective torque at \rin,
and increase the extent of the closed-field line region. The timescale
for magnetic diffusion is typically assumed to be about the same as
for viscous diffusion (which is supported by recent MRI calcuations;
\citealt{2009A&A...507...19F}), so that the radial diffusion of closed
field-lines outward could compete with accretion of matter (and field
lines) inward, creating an additional, longer timescale source of
variability in \delr. This could allow the instability to appear and
disappear at the same \mavg\ (as has been seen in
\citealt{2009ApJ...707.1296P}, where the 1-Hz QPO was seen in portions of
the outburst but not others at the same flux).

The physics that sets \del2r\ (the transition length between the
accreting and non-accreting solutions) is considerably less clear,
although presumably the same variations in the disc-field interaction
that change \delr\ could also change \del2r. In DS11 we discussed the
model proposed by \cite{2006ApJ...639..363P}, which considers a
strongly inclined dipole, so that the disc truncation radius lies both
inside and outside \rc, allowing for simultaneous accretion and
confinement of material. If this is the case then \del2r\ should
remain constant for a given system, and the variability must come from
fluctuations in \delr.

Another open question remains what regulates the mean accretion rate
onto the star. In all our simulations we have treated the accretion
rate, \mavg\ as a free parameter, and shown how the instability and
spin-evolution of the star change as a result. However, observations
of neutron star binaries suggest that the accretion rate from the
donor star remains roughly constant \cite{1997ApJS..113..367B}, and
variations in \mavg\ are often assumed to arise from ionization
instabilities \citep{2001NewAR..45..449L}. \cite{2010arXiv1006.1908H}
suggested that the disc-field interaction could regulate the accretion
rate onto the star by storing it up, in a similar way to what we have
discussed here. However, our model does not agree with this
conclusion: the disc-field interaction can only halt accretion long
enough to build up mass to drive the disc instability and only substantially
alters the inner regions of the disc. Our trapped disc solutions
evolve {\em as a response} to the decreased accretion rate in the
disc, and only become relevant once the average accretion rate has
decreased below a threshold, $\dot{m}_{\rm crit}/\dot{m}_{\rm c}
 \sim 4 \Delta r/r_{\rm in}$ (sec. \ref{sec:appearance}). 

On the other hand, if radiation feedback
from accretion onto the star plays an important role in regulating the
viscosity (and so accretion rate) in the disc, then temporarily
suppressing accretion onto the star might cause the level of
ionization to drop and drive the source back into quiescence.

A final question remains: why do some sources show the accretion
instability and trapped disc behaviour and not others? The vast
majority of outbursting neutron stars show long periods of quiescence,
consistent with being driven by the ionization instability with \rin\
far from \rc, rather than remaining trapped near \rc. Of the two
sources we have considered, IGR J00291+5934 has previously shown a normal
outburst pattern (a much stronger outburst followed by a long
quiescence; \citealt{2010arXiv1006.1908H}.

In their discovery paper for NGC 6440 X-2, \cite{2010ApJ...712L..58A}
remark that the short recurrence time and weak outburst likely cause
similar NS sources to be missed by current surveys, so it is possible
that such sources are common but remain undiscovered. As well, in
DS11 we showed that the very small ratio $T_{\rm visc}/T_{\rm SD}
\sim 10^{-17}$ for rapidly spinning pulsars made it more likely that
the disc would become untrapped (with \rin\ moving rapidly away from
\rc) as $\dot{m} \rightarrow 0$. \cite{2010arXiv1006.1908H} suggested
that the disc might remain weakly ionized during the entire outburst,
which would then decrease the viscosity and increase $T_{\rm visc}$,
perhaps enough to prevent the disc from becoming untrapped. If the RI
instability is responsible for the 1Hz QPO seen in NGC 6440 X-2 and
SAX J1808.4-3658, then the fact that only sometimes manifests is
consistent with the idea that \delr\ is variable (since the
instability only occurs over a small range in \delr ). 

For T~Tauri stars, the main constraint on the instability is the
presence of a strong ($\sim 1$ kG) dipolar magnetic field. Although
most T~Tauri stars show a strong surface field component, the dipolar
component can vary considerably, to the point that it is not
clear whether the disc-field interaction considered here is primarily
responsible for the low spin-rate of the star. The inclination
between the rotation axis of the star and the magnetic field could
also affect the instability by changing \del2r\ (as above). This
would suggest that, as a class, EXors are T~Tauri stars with moderate
accretion rates and strong, aligned dipolar magnetic fields compared
with normal T~Tauri stars.

\section{Conclusions}
\label{sec:conclusion}
\bhl As found in DS10, accretion from a disc onto a magnetosphere
tends to take place in what we have called `trapped' states. The inner
edge of the disc hovers around the corotation radius in this state,
even at very low accretion rates. The accretion in this state can be
steady or cyclic. We have investigated here the conditions under which
the cyclic form takes place, the kinds of cycles, their amplitudes,
cycle periods and their effect on the average torque on the accreting
star.

Two forms of cycle are found; the first form (labeled RI here) is the
accretion/dead disc cycle already suggested by
\cite{1977PAZh....3..262S}. A pile-up of mass at the `centrifugal
barrier' outside corotation is followed by an episode of accretion
emptying the pile and the beginning of a new cycle. Mass loss or
`propellering', whether it takes place or not, is a separate piece of
physics not needed for understanding the effect of a centrifugal
barrier on a viscous disc accreting on a magnetosphere.

Whether accretion is cyclic or continuous is found to depend on the width 
of the transition from an accreting state (inner edge inside corotation) to
a suspended accretion state (edge outside corotation). A narrow width leads
to RI cycles, a broader transition to continuous accretion. The effect of these
cycles is to increase the average spindown torque on the star. A second form of
cycle takes places when the average accretion rate is close to the standard
accreting case (edge inside corotation). It decreases the average torque
on the star. 

A review of the available literature shows forms of variability that
may be related the the cyclic forms of accretion found here, the most
promising ones may be the `EX Lup' outbursts, and a new form of cyclic
accretion found in X-ray binaries NGC 6440 X-2 and SAX J1808.4-3658.
\ehl

\section{Acknowledgments}
CD'A acknowledges financial support from the National Science and
Engineering Research Council of Canada, and thanks Attila Juhasz for
interesting dicussions and the use of figure \ref{fig:EXLup}.

\bibliographystyle{mn2e}
\bibliography{magbib}
\label{lastpage}
\end{document}